\documentclass[aps,reprint,nofootinbib,nobibnotes,notitlepage,superscriptaddress,onecolumn,prd,
 amsmath,amssymb
]{revtex4-2}

\usepackage[caption=false]{subfig}
\usepackage{braket}
\usepackage{float}
\usepackage{lipsum}
\usepackage{graphicx}
\usepackage{dcolumn}
\usepackage{bm}
\usepackage{natbib}
\usepackage{hyperref}
\hypersetup{
	colorlinks = true,
    linkcolor = Red,
    urlcolor  = Red,
    citecolor = Red
}

\usepackage{microtype}

\usepackage{verbatim}
\usepackage[amssymb]{SIunits}
\usepackage{tabularx}
\usepackage[dvipsnames]{xcolor}
\usepackage{wasysym}

\usepackage{tikz}

\def\be{\begin{equation}}
\def\ee{\end{equation}}

\usepackage{color}
\definecolor{darkgreen}{RGB}{0,120,0}

\definecolor{darkgreen}{RGB}{0,120,0}

\newcommand{\resub}[1]{{#1}}

\newcommand{\Mpch}{h^{-1}\mathrm{Mpc}}
\newcommand{\hMpc}{h\,\mathrm{Mpc}^{-1}}

\newcommand{\delD}[1]{(2\pi)^3\delta_\mathrm{D}\left({#1}\right)}

\newcommand{\av}[1]{\left\langle{#1}\right\rangle} 

\newcommand{\vk}{\vec k}
\newcommand{\hk}{\hat{\vec k}}

\newcommand{\vs}{\vec s}

\newcommand{\vx}{\vec x}
\newcommand{\hx}{\hat{\vec x}}

\newcommand{\hr}{\hat{\vec r}}
\newcommand{\hn}{\hat{\vec n}}

\newcommand{\hs}{\hat{\vec s}}
\newcommand{\tjo}[3]{\begin{pmatrix} {#1} & {#2} & {#3}\\ 0 & 0 & 0\end{pmatrix}}
\newcommand{\tj}[6]{\begin{pmatrix} {#1} & {#2} & {#3}\\ {#4} & {#5} & {#6}\end{pmatrix}}

\renewcommand{\vr}{\vec r}
\renewcommand{\L}{\Lambda}
\renewcommand{\P}{\mathcal{P}}

\def\beq{\begin{eqnarray}}
\def\eeq{\end{eqnarray}}
\def\k{\textbf{k}}
\def\bfk{\textbf{k}}
\def\bfs{\textbf{s}}
\def\bft{\textbf{t}}
\def\bfu{\textbf{u}}
\let\vec\mathbf

\makeatletter
\newlength{\apb@width}
\newcommand{\autoparbox}[2][c]{\settowidth{\apb@width}{#2}\parbox[#1]{\apb@width}{#2}}
\newcommand{\includegraphicsbox}[2][]{\autoparbox{\includegraphics[#1]{#2}}}
\makeatother

\def\mnras{\rm{MNRAS}}
\def\aj{\rm{AJ}}
\def\aap{\rm{A\&A}}

\def\physrep{\rm{Phys.~Rep.}}
\def\apjs{\rm{ApJS}}               
\def\jcap{\rm{JCAP}}

\begin{document}

\title{\LARGE Colliding Ghosts:\\ 
\large Constraining Inflation with the Parity-Odd Galaxy Four-Point Function}

\author{Giovanni Cabass}
\email{gcabass@ias.edu}
\affiliation{School of Natural Sciences, Institute for Advanced Study, 1 Einstein Drive, Princeton, NJ 08540, USA}

\author{Mikhail M.~Ivanov}
\email{ivanov@ias.edu}
\affiliation{School of Natural Sciences, Institute for Advanced Study, 1 Einstein Drive, Princeton, NJ 08540, USA}
\affiliation{NASA Hubble Fellowship Program Einstein Postdoctoral Fellow}

\author{Oliver~H.\,E.~Philcox}
\email{ohep2@cantab.ac.uk}
\affiliation{Center for Theoretical Physics, Department of Physics,
Columbia University, New York, NY 10027, USA}
\affiliation{Simons Society of Fellows, Simons Foundation, New York, NY 10010, USA}
\affiliation{Department of Astrophysical Sciences, Princeton University, Princeton, NJ 08540, USA}
\affiliation{School of Natural Sciences, Institute for Advanced Study, 1 Einstein Drive, Princeton, NJ 08540, USA}

\begin{abstract} 

    \noindent Could new physics break the mirror symmetry of the Universe? Utilizing recent measurements of the parity-odd four-point correlation function of BOSS galaxies, we probe the physics of inflation by placing constraints on the amplitude of a number of parity-violating models. Within canonical models of (single-field, slow-roll) inflation, no parity-asymmetry can occur; however, it has recently been shown that breaking of the standard assumptions can lead to parity violation within the Effective Field Theory of Inflation (EFTI). In particular, we consider the Ghost Condensate and Cosmological Collider scenarios -- the former for the leading and subleading operators in the EFTI and the latter for different values of mass and speed of an exchanged spin-$1$ particle -- for a total of $18$ models. Each instance yields a definite prediction for the inflationary trispectrum, which we convert to a late-time galaxy correlator prediction (through a highly non-trivial calculation) and constrain using the observed data. We find no evidence for inflationary parity-violation (with each of the $18$ models having significances below $2\sigma$), and place the first constraints on the relevant coupling strengths, at a level comparable with the theoretical perturbativity bounds. This is also the first time Cosmological Collider signatures have directly been searched for in observational data. We further show that possible secondary parity-violating signatures in galaxy clustering can be systematically described within the Effective Field Theory of Large-Scale Structure. We argue that these late-time contributions are subdominant compared to the primordial parity-odd signal for a vast region of parameter space. In summary, the results of this paper disfavor the notion that the recent hints of parity-violation observed in the distribution of galaxies are due to new physics. 
    
\end{abstract}

\maketitle

\section{Introduction}
\label{Sec:Intro}

\noindent Cosmic inflation probes physics at energy scales vastly above those of terrestrial experiments, providing a unique window into fundamental physics. Whilst the inflationary period cannot be observed directly, quantum fluctuations produced therein source perturbations in the metric, which manifest themselves in the distribution of matter and gravitational waves today. As such, careful analysis of late-time observables, such as the cosmic microwave background (CMB) and large-scale structure (LSS), can be used to shed light on primordial physics.

\begin{figure}
    \centering
    \includegraphics[width=0.5\textwidth]{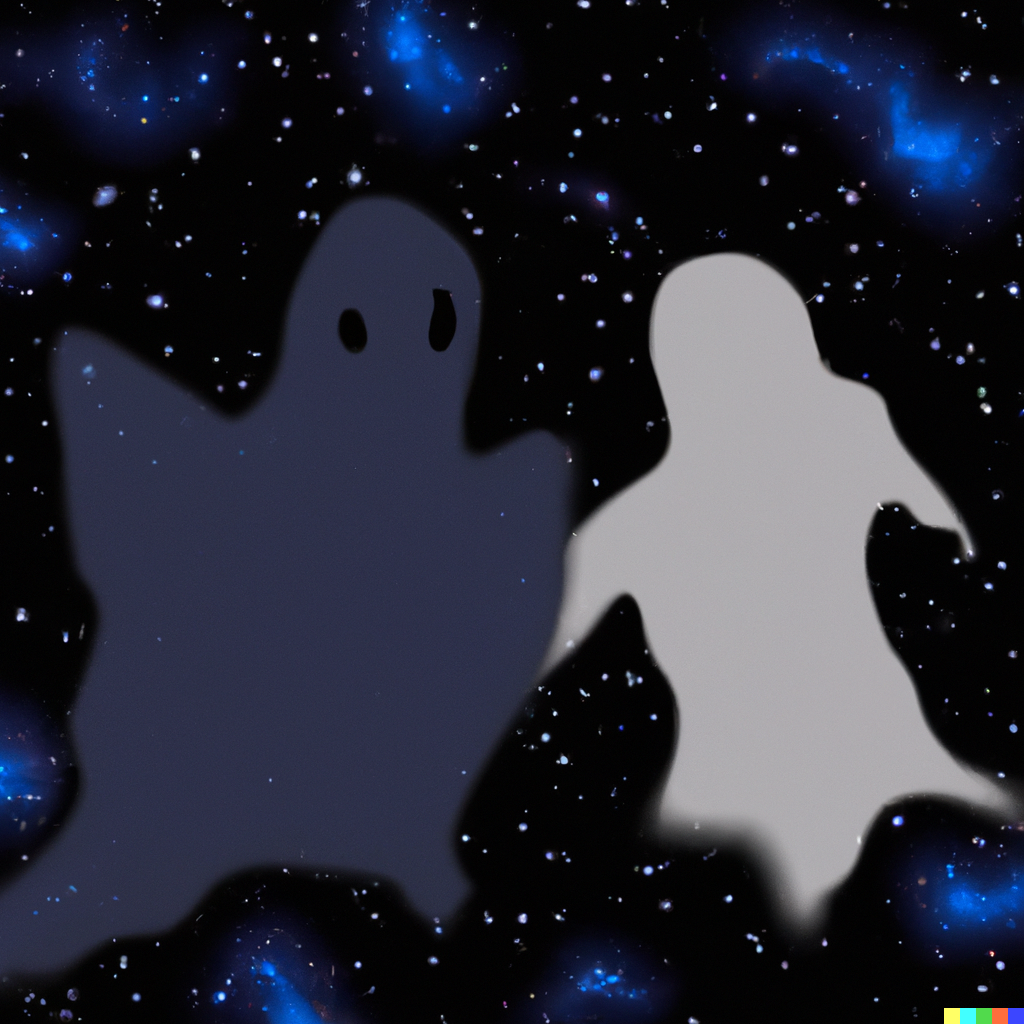}
    \caption{Outline of this work \citep{ramesh2021zero}.}
    \label{fig: schematic}
\end{figure}

The simplest models of inflation predict a Gaussian spectrum of primordial perturbations \citep{Guth:1980zm,Starobinsky:1980te,Linde:1981mu}, and thus a Gaussian distribution for the CMB (neglecting secondary effects), and LSS (within the linear regime). In this scenario, the early Universe is controlled by a single scalar field, known as the inflaton, whose dynamics are set by a quadratic action on a (nearly) de-Sitter background \citep[e.g.,][]{Mukhanov:1981xt,Starobinsky:1982ee}. A wide variety of extensions to this exist, involving, for example, additional fields (either massive or massless, and of varying spin \citep[e.g.,][]{Arkani-Hamed:2015bza}), modified kinetic terms (such as those motivated by UV completions involving extra dimensions \citep[e.g.,][]{Silverstein:2003hf}), and non-standard vacua. In recent years, a particularly useful framework to categorize this landscape has emerged, in the form of the Effective Field Theory of Inflation (hereafter EFTI) \citep{Cheung:2007st} (see \citep{Cabass:2022avo} for a recent review). In the vein of other effective field theories, this systematically predicts the various operators that can appear in the low-energy action, consistent with various symmetry principles. To probe inflation, we can thus search for the late-time signatures of these terms.

A generic prediction of many non-standard inflationary theories is a modification of the primordial spectrum of density perturbations, for example by introducing skewness to the initially Gaussian distribution (see \citep{Bartolo:2004if,Meerburg:2019qqi} for a review). In its simplest form, this sources a three-point correlator (or bispectrum) of the primordial curvature perturbation, the shape of which is fixed by the inflationary Lagrangian. By winding forward the cosmological clock, one can calculate the impact on late-time statistics, which takes the form of a non-trivial CMB and LSS three-point function. Since the early 2000s, various attempts have been made to constrain such models using both the CMB, and, more recently, spectroscopic surveys, usually by measuring a set of characteristic non-Gaussian amplitudes, $f_{\rm NL}$, which can be related to couplings in the EFTI \citep[e.g.,][]{2016A&A...594A..17P,2020A&A...641A..10P,Senatore:2009gt,Cabass:2022wjy,Cabass:2022ymb,DAmico:2022gki,Mueller:2021tqa,Alvarez:2014vva,Duivenvoorden:2019ses}. Whilst no detections have been yet made, bounds will continue to tighten with upcoming experiments such as the Simons Observatory \citep{2019JCAP...02..056A} and the Dark Energy Spectroscopic Instrument (DESI) \citep{2016arXiv161100036D}, and there is some hope of probing interesting regimes such as the $f_{\rm NL}\approx 1$ limit in the future (which is a natural boundary for many models of multi-field inflation).

The rich landscape of inflationary physics source much more than just primordial three-point functions. At next order, one may consider kurtosis, \textit{i.e.}\ the primordial four-point correlator (or trispectrum). This can be generated via a number of physical channels, such as particle exchange during inflation \citep[e.g.][]{Arkani-Hamed:2015bza,Bartolo:2010di}. As before, the EFTI predicts specific correlator templates with accompanying amplitudes. Some work has been performed to constrain these with the CMB (usually via the $g_{\rm NL}$ and $\tau_{\rm NL}$ amplitude parameters) \citep[e.g.,][]{2015arXiv150200635S,2020A&A...641A...9P,Vielva:2009jz,Sekiguchi:2013hza}, but the field is still in its infancy, and is hampered by the comparably low signal-to-noise of higher-point statistics in nature.

An intriguing feature of scalar four-point functions (such as the primordial curvature perturbation) is that they are \textit{chiral}, \textit{i.e.}\ one can define a handedness to the shapes which flips under mirror-reflection. 
This was first pointed out in the galaxy-survey context in \citep{2021arXiv211012004C} (see also \citep{1999PhRvL..83.1506L,2010PhRvD..81l3529G,2020JHEP...04..189L}).
If the primordial Universe preserves parity symmetry, there should be no difference between left- and right-handed shapes, thus the parity-odd part of the four-point function should vanish. In the late Universe, large-scale physics is set by gravity and hydrodynamics, which (at least in conventional theories) conserve parity, but interesting violations could occur during inflation. Indeed, the creation of the known baryon-antibaryon imbalance requires some form of primordial charge-parity asymmetry \citep[e.g.,][]{1967JETPL...5...24S,2004PhRvL..93t1301D,2013JCAP...04..046A,2016IJMPD..2540013A}. 

To understand inflationary parity-violation, we can once again look to the EFTI. Assuming a scale-invariant Universe with a Bunch-Davies vacuum, populated by a set of arbitrary scalar fields (with interactions that fall off sufficiently fast as the modes are stretched outside the horizon during inflation), the parity-odd primordial trispectrum vanishes at tree level \citep{2020JHEP...04..189L,Cabass:2022rhr}. Any significant detection would thus indicate violation of one of the above assumptions, and could hint at a variety of non-standard inflationary scenarios. Intriguing examples of this include non-standard vacua such as ghost condensation \citep{Arkani-Hamed:2003juy,Arkani-Hamed:2003pdi}, (strong) violation of scale invariance, or the exchange of massive spinning particles \citep{2020JHEP...04..189L,Cabass:2022rhr}. As for the bispectrum, each scenario arises from specific terms in the EFTI Lagrangian, whose signatures can be searched for in late-time observables. 

If we wish to probe the inflationary peculiarities described above, we require parity-sensitive observables. In general, late-time observables fall into two categories: those sensitive to tensor perturbations (\textit{i.e.}\  gravitational waves) and those sensitive to scalar perturbations (\textit{i.e.}\ gravitational potentials). Quantities in the first class include CMB polarization \resub{(including $V$ modes) \citep[e.g.,][]{2011PhRvD..83b7301K,2008PhLB..660..444A,2017PhRvL.118v1301M,2017JCAP...07..034B,Planck:2016soo,Gerbino:2016mqb,2015JCAP...01..027B,2015JCAP...07..039B,2018PhRvD..98d3533F,Bartolo:2018elp,Orlando:2022rih}}, galaxy shapes \resub{\citep{Biagetti:2020lpx}}, galaxy spins \citep{Yu:2019bsd,Motloch:2021mfz}, \resub{and directly-observed stochastic gravitational waves \citep{Orlando:2020oko}}. Through their dependence on chiral gravitational waves, these can have parity-sensitive power spectra, for instance, the CMB $TB$ correlator \resub{\citep{Bordin:2020eui,Bartolo:2020gsh,Cabass:2022jda}}. In contrast, the second class of observables, including CMB $T$- and $E$-modes and LSS density fields, depend only on inflationary scalars, thus parity-sensitivity appears only in the trispectrum and beyond (since a parity transformation is equivalent to a rotation for the power spectrum and bispectrum) \citep{2016PhRvD..94h3503S,Cabass:2022rhr,1999PhRvL..83.1506L,2014JCAP...12..050D}. In this work, we consider the latter case, noting that the physical origins of scalar- and tensor-type parity violation can be distinct, and there has been no robust detection of gravitational wave signatures to date.

In this work, our primary observable is the galaxy four-point correlation function, which is the configuration-space analog of the galaxy trispectrum. The parity-even and parity-odd contributions to this 
(as defined in \citep{2021arXiv211012004C}) 
were measured for the BOSS galaxy survey in \citep{2021arXiv210801670P} and \citep{Philcox:2022hkh,Hou:2022wfj} respectively, and, intriguingly, there are some hints of a non-zero signal in the latter, roughly at the $3\sigma$ level. Na\"ively interpreted, this could be a smoking gun of non-standard inflationary physics, sourced by models such as those considered above. We caution that this is not the only possible explanation. Though conventional post-reheating physics is thought to be parity-conserving (at least on the large scales relevant to galaxy clustering, $r\geq 20\Mpch$ here), more esoteric suggestions such as Chern-Simons modified gravity \citep{2009PhR...480....1A} could lead to a late-time signal. Furthermore, the measurements themselves are wrought with complexity, as is their interpretation. Little work has been devoted to the impact of data systematics (such as window functions and galactic dust) on the higher-point functions, and modeling the noise properties of the data is no mean feat. In the latter case, knowledge of the (connected) eight-point function is strictly required, and the detection significance varies wildly with different approaches to its estimation \citep{Philcox:2022hkh,Hou:2022wfj}. 

Potential hints of cosmic parity-breaking motivate careful study of its possible origins. In \citep{Philcox:2022hkh}, a single inflationary model was considered, involving a non-decaying $U(1)$ gauge field coupled to the inflaton via a Chern-Simons interaction, as proposed in \citep{2015JCAP...07..039B,2016PhRvD..94h3503S}. This has a number of theoretical problems, in particular, the (admittedly small) anisotropy imprinted in the two-point function, and the inherent tachyonic instability, which leads to the exponential amplification of higher-point functions. Indeed, no evidence was found for this model in the former work. Here, we consider two types of parity-breaking motivated by the EFTI and for which the theoretical prediction is under perturbative control: Ghost Inflation \citep{Arkani-Hamed:2003juy} and the exchange of a spin-$1$ particle (part of the Cosmological Collider setup, \citep[e.g.,][]{Arkani-Hamed:2015bza,Chen:2009zp,Baumann:2011nk}), following the derivations of \citep{Cabass:2022rhr}. In particular, we will place constraints on the corresponding inflationary couplings, and additionally assess whether these scenarios could be responsible for the observed parity-excess. Further, we will compare our constraints to the rough magnitudes expected from the EFTI via perturbativity bounds, allowing us to assess whether such our constraints are parametrically relevant. This paper provides the first bounds on such models of inflation; however, the constraining power will only grow in the future with the advent of new surveys and new datasets. 

Whilst not the main focus of this paper, it is interesting to note that modern EFT techniques in cosmology -- namely the Effective Field Theory of Large-Scale Structure (EFTofLSS; \citep{Baumann:2010tm,Carrasco:2012cv}, see \citep{Cabass:2022avo} for a recent review and \citep{2020JCAP...05..042I,2020JCAP...05..032P,2020JCAP...05..005D,Chen:2021wdi} for applications to data) -- allow one to parameterize any \textit{late-time} sources of parity-violation in a model-independent way, giving rise to templates that are compatible with all the symmetries of the large-scale structure barring point reflection \resub{(assuming the equivalence principle)}. The measurement of \citep{Philcox:2022hkh,Hou:2022wfj} could be used to put constraints on the amplitude of these templates, which may then be translated to bounds on the ``microphysical'' parameters of any late-time model. The power of such EFT techniques is that they allow us, again in a model-independent manner, to estimate the size of parity violation. The final result of this work is to show that these would-be signatures are much smaller than the parity-even contributions from gravitational collapse to the galaxy four-point function if the spatial nonlocality scale associated with them is of the same order as the non-linear scale of structure formation. With the same assumptions, we also show that they would be subdominant to primordial contributions if we are in the regime of mild primordial non-Gaussianity. Again, we stress that these conclusions hold regardless of the fact that these signatures would require some form of parity-violating gravity (or hydrodynamics) operating at late times. In summary, the results of this paper \resub{yield three possible explanations for the results of \citep{Philcox:2022hkh,Hou:2022wfj}: (1) inflationary physics uncorrelated with the models tested herein; (2) late-time physics with a huge correlation length; (3) systematics in the data or analysis procedure.} 

\vskip 4pt
The remainder of this paper is structured as follows. In \S\ref{sec:early_universe}, we discuss parity-violating inflation, and introduce the models considered in this work and their corresponding primordial trispectra. \S\ref{sec: data} discusses the observable utilized herein (the galaxy four-point correlation function), before we present theoretical predictions for its form in \S\ref{sec: transform}. Our main results, including amplitude constraints, are given in \S\ref{sec: results}. \S\ref{sec: late-time} discusses the parametrization of late-time parity violation within the EFTofLSS, before we conclude in \S\ref{sec: conclusion}. Gory details of the calculations are presented in Appendix~\ref{app: calculations}.

\section{Inflationary Parity Violation} 
\label{sec:early_universe}

\noindent In this section we introduce two candidate models for parity-violating inflation: the Ghost Condensate \cite{Arkani-Hamed:2003juy,Arkani-Hamed:2003pdi} and the Cosmological Collider \cite{Bordin:2018pca,CosmoBootstrap1,CosmoBootstrap2,Baumann:2022jpr,Arkani-Hamed:2015bza,COT,MLT,BBBB,2020JHEP...04..189L,Qin:2022fbv,Baumann:2011nk,Chen:2009zp}. This follows from \citep{Cabass:2022rhr}, which studied these two scenarios as examples of models that evade general theorems about parity violation in the scalar sector, and can give rise to a parity-odd trispectrum for the comoving curvature perturbation $\zeta$.

The first case can be seen as a limit of the Effective Field Theory of Inflation (EFTI) \citep[e.g.,][]{Cheung:2007st,Cabass:2022avo} in which the quantum fluctuations, $\pi$, of the clock (which on superhorizon scales are simply proportional to $\zeta$) have a dispersion relation $\omega^2\propto k^4$. An example of UV completion is a scalar field $\phi$ with a Lagrangian that is a function $P$ of $\smash{X\equiv{-(\partial_\mu\phi)^2}}$ such that excitations $\pi$ about the ``trivial'' background $\phi=0$ are unstable, but those around the background $\phi = \mu t$ are not. If ${\rm d}P/{\rm d}X$ vanishes on the background, $\pi$ will have a nonrelativistic dispersion relation \cite{Arkani-Hamed:2003pdi,Arkani-Hamed:2003juy}. Note that ghost 
condensation naturally 
arises as a low energy limit of models
with Lorentz invariance violation
in the inflaton sector~\cite{Ivanov:2014yla}.

In the second instance, one considers the impact of massive spinning particles, $\sigma^{ij\cdots}$, coupled to the clock. Even if these particles decay on superhorizon scales, they can be created from the vacuum and exchanged by $\pi$ fluctuations in the bulk of de Sitter spacetime, and leave an impact on the statistics of the curvature perturbation $\zeta$ that are not degenerate with local operators in the EFTI if their mass,  $m_\sigma$, is comparable to the Hubble scale, $H$. 

Below, we briefly recapitulate the interactions studied in \citep{Cabass:2022rhr}, summarize the corresponding templates for the parity-odd trispectrum (which will be used to predict the parity-odd galaxy correlator in \S\ref{sec: transform}), and discuss bounds on their size from requirements of perturbativity. \resub{In all cases, we assume the standard EFTI symmetries, and work in the close-to-de-Sitter limit, in which templates are scale invariant, with any deviations slow-roll suppressed.}

\subsection{The Inflationary Lagrangian and Inflaton Interactions}

\noindent In the case of Ghost Inflation, the primordial Universe is described by a single clock $\pi$ (hereafter known as the Goldstone mode), which obeys the quadratic action 
\beq
\label{eq:leading_action} 
S_{\pi\pi}=\int{\rm d}^4x\,\sqrt{-g}\,\bigg[\frac{\Lambda^4}{2}{\dot{\pi}^2} - \frac{\tilde{\Lambda}^2}{2}\frac{(\partial^2\pi)^2}{a^4}\bigg]\,\,, 
\eeq
where $a$ is the scale factor, and the scales $\Lambda$ and $\tilde{\Lambda}$ control the normalization of the power spectrum (using $\zeta={-H\pi}$ on superhorizon scales):
\beq
\label{eq:Delta2zeta_GI}
k^3P_\zeta(k)\equiv\Delta^2_\zeta = \frac{H^2(H\tilde{\Lambda})^{\frac{1}{2}}\Gamma(\frac{3}{4})^2}{\pi\Lambda\tilde{\Lambda}^2} 
\eeq
for a scale-invariant power spectrum $\Delta^2_\zeta$. On subhorizon scales, $\pi$ follows the dispersion relation $\omega = \tilde{\Lambda}k^2/\Lambda^2$; it is this non-linear relation that results in the different phenomenology of the theory to standard single-field inflation. 

At tree level, the only contribution to a parity-odd trispectrum of the Goldstone mode (\textit{i.e.}\ the part of $\av{\pi\pi\pi\pi}$ antisymmetric under reflections) can come from contact diagrams. \citep{Cabass:2022rhr} studied the following two interactions, appearing at leading- and subleading-order in the effective field theory expansion respectively
\begin{align}
S^{(\rm LO)}_{\pi\pi\pi\pi} &= \frac{1}{M_{\rm PO}} \int {\rm d}^4 x\,\sqrt{-g}\,a^{-9}  \epsilon_{ijk}\partial_m\partial_n \pi \partial_n\partial_i {\pi}\partial_m\partial_l \partial_j {\pi} \partial_l \partial_k{\pi}\,\,, \label{gi_op-1} \\
S^{(\rm NLO)}_{\pi\pi\pi\pi} &= \frac{1}{\Lambda_{\rm PO}^2} \int {\rm d}^4x\, \sqrt{-g}\,a^{-9}\dot{\pi}\epsilon_{ijk}\partial_i\partial_l\pi\partial_l\partial_j\partial^2{\pi}\partial_k\partial^2{\pi}\,\,, \label{gi_op-2} 
\end{align}
where $\epsilon_{ijk}$ is the antisymmetric tensor, which gives rise to the parity-violation. It is important to keep in mind that these two operators fully exhaust only the subset of quartic operators that in the flat-space limit of the EFTI are invariant under the non-linear part $\delta\pi = \lambda_i x^i$ of the spontaneously broken Lorentz boosts. A full classification including Wess-Zumino terms is left for future work: in this analysis we focus on \eqref{gi_op-1} and \eqref{gi_op-2} as the simplest trispectrum-inducing couplings that arise due to deviations from a Bunch-Davies vacuum with a linear dispersion relation. 

Regarding the Cosmological Collider, in this work we focus on the same setup studied in \cite{Cabass:2022rhr}, \textit{i.e.}\ the parity-odd four-point function arising from the exchange of a massive spin-$1$ field $\sigma^i$. This has the Feynman diagram
\begin{equation}
\label{eq:exchange_diagram}
\raisebox{-0.0cm}{\includegraphicsbox[scale=1.3]{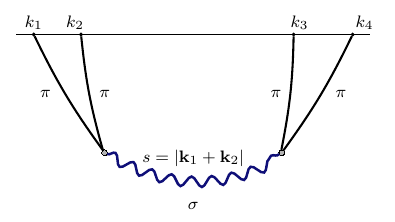}} 
\end{equation}
for the $s$-channel exchange. 
The parity-even and parity-odd vertices are, respectively: 
\begin{align} 
S^{\rm PE}_{\pi \pi \sigma} &= \lambda_{1}\int {\rm d}^4 x\,\sqrt{-g}\,a^{-3}\partial_i \dot{\pi} \partial_i \partial_j \pi \sigma^j\,\,, \label{CubicAction-1} \\ 
S^{\rm PO}_{\pi \pi \sigma} &= \lambda_{3}\int {\rm d}^4 x\,\sqrt{-g}\,a^{-4}\epsilon_{ijk} \partial_i \partial_l \pi \partial_j \partial_l \dot{\pi} \sigma^k\,\,, \label{CubicAction-2}
\end{align}
where $\lambda_{1}$ has dimensions of energy and $\lambda_{3}$ is dimensionless.\footnote{Notice that here we assume that $\sigma^i$ transforms as a vector under parity.} 
The quadratic actions for $\pi$ and $\sigma^i$, instead, are given by 
\begin{align} 
    \label{sigmafree}
    S_{\pi\pi} &=\frac{H^4}{4c^3_s\Delta^2_\zeta}\int {\rm d}^4x\,\sqrt{-g}\left[\dot{\pi}^2 - c^2_sa^{-2}(\partial_i\pi)^2\right]\,\,, \\
    S_{\sigma\sigma} &=\frac{1}{2}\int {\rm d}^4x\,\sqrt{-g} \left[(\dot{\sigma}^i)^2- c_1^2a^{-2} (\partial_i \sigma^j)^2 -(c_0^2-c_1^2)a^{-2} (\partial_i\sigma^i)^2 - m_{\sigma}^2(\sigma^i)^2 \right]\,\,,
\end{align}
where all indices are raised and lowered with $\delta_{ij}$, $c_{0,1}$ are the speeds of sound of the longitudinal and transverse components of $\sigma^i$ and $m_{\sigma}$ is the mass of the spin-1 field. In the exchange diagram of \eqref{eq:exchange_diagram} only the $\pm 1$ helicities are exchanged. Without loss of generality we can then set $c_1=1$, so that $c_s > 1$ ($c_s < 1$) means that the $\pi$ fluctuations are moving faster (slower) than the spinning particle. 

It is important to keep in mind that the operators of \eqref{CubicAction-1} and \eqref{CubicAction-2} do not exhaust all the possible signatures of parity violation in the Cosmological Collider:
\begin{itemize}
    \item here we are considering only operators that the non-linear realization of boosts does not tie to quadratic mixings between $\pi$ and the helicity-$0$ mode of $\sigma^i$. The space of interactions between two $\pi$ fluctuations and one $\sigma^i$ is larger than this;
    \item even with this restriction there are two other operators at leading order in the effective field theory expansion: $\smash{\lambda_2a^{-1}\ddot{\pi} \partial_i \dot{\pi} \sigma^i}$ and $\smash{\lambda_4a^{-2} \epsilon_{ijk} \partial_i \ddot{\pi} \partial_j \dot{\pi} \sigma^k}$. Hence the space of signatures of parity violation involves the $\lambda_1\lambda_3$, $\lambda_2\lambda_3$, $\lambda_1\lambda_4$ and $\lambda_2\lambda_4$ exchanges, in principle; 
    \item parity violation can also arise if there is a split of the $\pm1$ helicities of $\sigma^i$ exchanged in the diagram of \eqref{eq:exchange_diagram} \cite{2020JHEP...04..189L}. This can happen for example via the dimension-$3$ operator $a^{-1}\epsilon_{ijk}\sigma^i\partial_j\sigma^k$ at leading order in the effective field theory expansion (or, in a UV-complete framework, a Chern-Simons-like interaction, \citep[e.g.][]{1999PhRvL..83.1506L}). 
\end{itemize}
As discussed in the introduction, the purpose of this work is to show how already with current surveys we can put constraints on the presence and interactions of massive particles during inflation. Since we expect that BOSS data will not give parametrically different constraints if we consider the scenarios in the three bullet points above (for example if we had considered the $\lambda_2\lambda_3$, $\lambda_1\lambda_4$ or $\lambda_2\lambda_4$ exchanges), focusing on this example will suffice for our purpose until data from future surveys like DESI and Euclid are available (and indeed the various templates are likely highly correlated). Another reason why we focus on the $\lambda_1\lambda_3$ exchange is that it is only interaction the among the four that has been explicitly bootstrapped in \citep{Cabass:2022avo}, leading to a closed-form expression for the resulting $\zeta$ trispectrum. \resub{We further note that other templates are possible if one loosens the symmetries imposed on the EFTI, e.g., allows for strong departures from scale invariance.}

In this regard, let us also discuss our choice of spin and mass of the exchanged particle. Spin-$s$ particles must satisfy the Higuchi bound $m_\sigma^2>s(s-1)H^2$ \cite{Higuchi:1986py}, implying that the signature in $\zeta$ correlators of very massive particles with high spin that is not degenerate with EFTI operators is exponentially suppressed (in addition to the suppression in the squeezed limit due to their fast decay as the universe expands) \cite{Arkani-Hamed:2015bza}. In this work we focus on spin-$1$ particles since, in this case, any mass is allowed. It would be interesting to study the signature of higher-spin particles in the setup of \citep{Bordin:2018pca}: there the authors invoke strong couplings with the clock in order to evade the Higuchi bound, turning the Cosmological Collider into ``Cosmological Condensed Matter''.\footnote{See also \citep{Deser:2003gw,Baumann:2017jvh,Franciolini:2017ktv,Kehagias:2017cym,Maleknejad:2011sq,Maleknejad:2011jw,Adshead:2013nka,Agrawal:2017awz,Piazza:2017bsd} for other ways of having light spinning particles during inflation (more precisely particles belonging to unitary representations of the de Sitter group different than the ``principal series'' $m_\sigma^2\geq(s-1/2)^2$, a strong breaking of the shift symmetry of the clock, and symmetry breaking patterns different than that of the EFTI).} In keeping with the exploratory nature of our paper, we leave also this to future work. 

\subsection{Trispectrum templates}
\label{subsec:trispectrum_templates}

\noindent We now summarize the trispectrum templates of the above models, \textit{i.e.}\ the predictions for 
\begin{equation}
\label{eq: Tk-def}
    \av{\zeta(\vk_1)\zeta(\vk_2)\zeta(\vk_3)\zeta(\vk_4)}_c \equiv \delD{\vk_{1234}}T(\vk_1,\vk_2,\vk_3,\vk_4)\equiv\delD{\vk_{1234}}\tilde{T}(\vk_1,\vk_2,\vk_3,\vk_4)+\text{23 perms.}\,\,,
\end{equation}
where the second definition is explicitly symmetrized. Before doing so, we emphasize that we follow the recent inflationary analyses of BOSS data \citep{Cabass:2022wjy,Cabass:2022ymb,DAmico:2022gki} and consider only \emph{scale-invariant} templates for primordial non-Gaussianities.

\subsubsection{Ghost Condensate}

\noindent The pre-symmetrized trispectra from the operators of \eqref{gi_op-1} and \eqref{gi_op-2}, which we will denote with a subscript $M_{\rm PO}$ and $\Lambda^2_{\rm PO}$ respectively, are given by
\begin{equation}
\label{eq: primordial-ghost}
\begin{split}
\tilde{T}_{M_{\rm PO}}(\bfk_1,\bfk_2,\bfk_3,\bfk_4) &= {\frac{128i\pi^3 \Lambda^5(H\tilde{\Lambda})^{1/2}}{M_{\rm PO} \tilde{\Lambda}^5 \Gamma(\frac{3}{4})^2}}(\Delta^2_\zeta)^3\frac{(\k_1\cdot\k_2\times \k_3)(\k_2\cdot \k_4)(\k_1\cdot \k_4)(\k_2\cdot \k_3)}{k_1^{\frac{3}{2}}k_2^{\frac{3}{2}}k_3^{\frac{3}{2}}k_4^{\frac{3}{2}}}\,{\rm Im}\,{\cal T}^{(11)}_{0,0,0,0}(k_1,k_2,k_3,k_4)\,\,, \\ 
\tilde{T}_{\Lambda^2_{\rm PO}}(\bfk_1,\bfk_2,\bfk_3,\bfk_4) &= {\frac{512i\pi^3 \Lambda^5(H\tilde{\Lambda})^{3/2}}{\Lambda_{\rm PO}^2 \tilde{\Lambda}^6 \Gamma(\frac{3}{4})^2}}(\Delta^2_\zeta)^3(\k_1\cdot\k_2\times \k_3)(\k_1\cdot \k_2) k_1^{-\frac{3}{2}}k_2^{\frac{1}{2}}k_3^{\frac{1}{2}}k_4^{\frac{1}{2}}\,{\cal T}^{(13)}_{0,0,0,1}(k_1,k_2,k_3,k_4)\,\,, 
\end{split}
\end{equation}
where the function $\cal{T}$ is defined as 
\beq\label{eq: ghostT}
    \smash{{\cal T}^{(n)}_{\nu_1,\nu_2,\nu_3,\nu_4}(k_1,k_2,k_3,k_4)}  = \int_{0}^{+\infty}{\rm d}\lambda\, \lambda^{n}\,H^{(1)}_{{\frac{3}{4}-\nu_1}}(2ik^2_1\lambda^2)H^{(1)}_{{\frac{3}{4}-\nu_2}}(2ik^2_2\lambda^2)H^{(1)}_{{\frac{3}{4}-\nu_3}}(2ik^2_3\lambda^2)H^{(1)}_{{\frac{3}{4}-\nu_4}}(2ik^2_4\lambda^2)\,\,. 
\eeq 
The Hankel functions in the two integrals are exponentially convergent at large $\lambda$. 
We notice that $\smash{{\cal T}^{(n)}_{0,0,0,0}(k_1,k_2,k_3,k_4)}$ is purely imaginary and $\smash{{\cal T}^{(n)}_{0,0,0,1}(k_1,k_2,k_3,k_4)}$ is purely real, in keeping with the imaginary nature of parity-odd trispectra. We also see that there is no dependence of the trispectra on the Mandelstam-like variables 
\begin{align} 
\label{stu} 
\bfs &= \bfk_{1}+\bfk_{2}\,\,, & \bft &= \bfk_{1}+\bfk_{3}\,\,, & \bfu &= \bfk_{2}+\bfk_{3} \,\,, \\\nonumber
s &= |\bfk_{1}+\bfk_{2}|\,\,,&t& = |\bfk_{1}+\bfk_{3}|\,\,, & u& = |\bfk_{2}+\bfk_{3}|\,\,, 
\end{align} 
given that the trispectrum arises from contact diagrams (\textit{i.e.}\ without particle exchange). Finally, we notice that both templates contain the ubiquitous $\k_1\cdot\k_2\times \k_3$ factor, which is the only parity-violating structure possible for scale- and rotation-invariant trispectra. 

\subsubsection{Cosmological Collider}

\noindent To parametrize the mass of the spin-$1$ particle $\sigma$, we introduce the variable 
\beq
\nu\equiv\sqrt{\frac{9}{4}-\frac{m^2_\sigma}{H^2}}\,\,,
\eeq
which is real for $m_\sigma\leq 3H/2$. In this case, the trispectrum arising from the diagram of \eqref{eq:exchange_diagram} \resub{can be bootstrapped for arbitrary $c_s$ using the tools developed in \citep{Jazayeri:2022kjy}} and is given by 
\begin{equation}
\label{FactorisedTrispectrumCompact_zeta}
\begin{split}
 T_{\lambda_1\lambda_3}(\bfk_1,\bfk_2,\bfk_3,\bfk_4) &= {\left[\prod_{a=1}^4P_\zeta(k_a) \right]} \frac{c_{s}^4\lambda_{1}\lambda_{3}}{2H^3}\sin\pi\left(\nu+\frac{1}{2}\right)(s^2-k_{1}^2-k_{2}^2)(s^2-k_{3}^2-k_{4}^2)(k_{1}-k_{2})(k_{3}-k_{4}) \\ 
 &\;\;\;\;\;\;\,\,\times(\k_3\cdot\k_2\times\k_4) [k_{12}I_{3}(c_{s}k_{12},s)+ i c_{s}k_{1}k_{2}I_{4}(c_{s} k_{12},s)][k_{34}I_{4}(c_{s}k_{34},s)+ i c_{s}k_{3}k_{4}I_{5}(c_{s}k_{34},s)] \\
 &\;\;\;\;+[ (1,2) \leftrightarrow (3,4)] + t + u\,\,, 
\end{split}
\end{equation}
where $P_\zeta(k) =\Delta^2_\zeta/k^3$, we have defined $k_{ab}\equiv k_a+k_b$. The $6$ permutations in the above trispectrum are
\beq
\{1,2,3,4,1\!+\!2\}+\{3,4,1,2,3\!+\!4\}+
\{1,3,2,4,1\!+\!3\}+\{2,4,1,3,2\!+\!4\}+
\{1,4,2,3,2\!+\!3\}+\{2,3,1,4,1\!+\!4\}\,\,,
\eeq
and we require the functions $I_n$, defined as 
\begin{align}
\label{eq:nima_integral}
I_{n}(a,b) = (-1)^{n+1} \frac{H}{\sqrt{2 b}} \left(\frac{i}{2b} \right)^{n} \frac{\Gamma \left(\alpha \right)\Gamma \left(\beta \right)}{\Gamma(1+n)} \times {}_2 F_1 \Big(\alpha, \beta; 1+n; \frac{1}{2}-\frac{a}{2b}\Big) 
\end{align}
with $\smash{\alpha=\frac{1}{2}+n-\nu}$ and $\smash{\beta =\frac{1}{2}+n + \nu}$. These are such that $T_{\lambda_1\lambda_3}$ is purely imaginary, as required for a parity-odd trispectrum. Notably, the trigonometric prefactor vanishes if $m_\sigma=0$ or $m_\sigma=\sqrt{2}H$, \textit{i.e.}\ there is no contribution if the exchanged particle is massless or conformally coupled. 

For later use, we re-express \eqref{FactorisedTrispectrumCompact_zeta} in fully symmetrized form:
\begin{equation}
    \label{eq: primordial-collider-oliver}
    \begin{split}
    \tilde{T}_{\lambda_1\lambda_3}(\vk_1,\vk_2,\vk_3,\vk_4)&={-ic_{s}^4}\frac{\lambda_{1}\lambda_{3}}{\resub{2}H}(\Delta_\zeta^2)^4 \sin\pi\left(\nu+\frac{1}{2}\right)k_1^{-2}k_2^{-1}k_3^{-1}k_4^{-1}(\hk_1\cdot\hk_2)(\hk_3\cdot\hk_4)(k_{1}-k_{2})(k_{3}-k_{4}) \\ 
    &\;\;\;\;\times (\hk_2\cdot\hk_3\times\hk_4)[k_{12}J_{3}(c_{s}k_{12},s)+ c_{s}k_{1}k_{2}J_{4}(c_{s} k_{12},s)][k_{34}J_{4}(c_{s}k_{34},s)+c_{s}k_{3}k_{4}J_{5}(c_{s}k_{34},s)]\,\,, 
    \end{split}
    \end{equation}
noting that $T_{\lambda_1\lambda_3}$ is fully symmetric by definition, and introducing the rescaled integrals
\beq
J_{n}(a,b) = \left(\frac{1}{2b} \right)^{n+1/2} \frac{\Gamma \left(\alpha \right)\Gamma \left(\beta \right)}{\Gamma(1+n)} \,\, {}_2 F_1 \Big(\alpha, \beta; 1+n; \frac{1}{2}-\frac{a}{2b}\Big)\,\,.
\eeq

In the following we will restrict our analysis to the range of masses $0\leq m_\sigma\leq 3H/2$, though it would be interesting to extend the study of \citep{Cabass:2022rhr} to masses above $3H/2$ and see whether BOSS data are sensitive to the resulting oscillatory features in the trispectrum. 
We will also consider three choices for the (relative) speed of sound: $c_s=0.1,1,10$. At fixed $\lambda_1$ and $\lambda_3$, the overall size of the trispectrum scales as $1/c_s$ for $c_s\gg1$, and $c_s^4$ for $c_s\ll1$, thus we expect our constraints to be the strongest in the case where the Goldstone mode $\pi$ and the spinning particle move approximately with the same speed.

\subsection{Perturbativity Bounds}\label{subsec: perturbativity}

\noindent \textit{A priori}, it is not certain whether the BOSS data considered below will be able to place strong constraints on the above coupling amplitudes, nor what one even means by ``strong''. In this light, it is worth discussing the theoretical bounds on our trispectra that come from perturbativity, \textit{i.e.}\ the constraint that perturbation theory during inflation should be convergent.

\subsubsection{Ghost Condensate}

\noindent To enforce perturbativity, we require that the probability distribution functional of $\zeta$ is only weakly non-Gaussian.\footnote{An equivalent way to derive these constraints is requiring that at horizon crossing we are below the EFT cutoff as determined by the two operators \eqref{gi_op-1}\,\&\,\eqref{gi_op-2}.} 
We estimate the overall size, which we will call $\tau_{\rm NL}$, of the kurtosis as
\beq
\tau_{\rm NL} \sim \frac{T}{P_\zeta^3}\,\,. 
\eeq 
The requirement of perturbative non-Gaussianities can then be recast as \citep[e.g.,][]{Cabass:2022rhr}
\beq
\tau_{\rm NL}\Delta^2_\zeta\lesssim 1\,\,. 
\eeq

In these definitions we are cavalier with overall numerical factors and the shape dependence: forgetting the former will not lead to a parametric underestimation or overestimation of $\tau_{\rm NL}$, and we do not expect wild variations in the size of the trispectrum depending on the shape of the tetrahedron. From \eqref{gi_op-1}\,\&\,\eqref{gi_op-2} we then find 
\begin{align}
    \label{tauNL_in_Hubble}
    \tau_{\rm NL}^{(M_{\rm PO})} 
    \sim\frac{\Lambda^5H^{1/2}}{M_{\rm PO} \tilde{\Lambda}^{9/2}}\,\,, 
    \qquad 
    \tau_{\rm NL}^{(\Lambda^2_{\rm PO})} 
    \sim\frac{\Lambda^5H^{3/2}}{\Lambda_{\rm PO}^{2}\tilde{\Lambda}^{9/2}}\,\,. 
\end{align} 
In \S\ref{sec: results} we will put bounds on the combinations on the right-hand sides of these equations, and check whether BOSS data have a constraining power parametrically close to the perturbativity bound 
\begin{align}
    \label{tauNL_bounds}
    \tau_{\rm NL}^{(M_{\rm PO})}\lesssim 10^{8}\,\,, 
    \qquad 
    \tau_{\rm NL}^{(\Lambda^2_{\rm PO})}\lesssim 10^{8}\,\,. 
\end{align} 

Before proceeding, it is interesting to point out that using \eqref{eq:Delta2zeta_GI} we can express the Hubble rate in terms of the measured amplitude of the power spectrum, and find an expression for $\smash{\tau_{\rm NL}}$ only in terms of the ``microphysical'' scales $\smash{\Lambda}$, $\smash{\tilde{\Lambda}}$, $\smash{M_{\rm PO}}$ and $\smash{\Lambda^2_{\rm PO}}$. More precisely we find the scalings
\begin{align}
\label{removed_H_GI}
    \tau_{\rm NL}^{(M_{\rm PO})} 
    \sim\frac{1}{M_{\rm PO}}\sqrt[5]{\frac{\Lambda ^{26}}{\tilde{\Lambda}^{21}}} \,\,, 
    \qquad 
    \tau_{\rm NL}^{(\Lambda^2_{\rm PO})} 
    \sim\frac{1}{\Lambda^2_{\rm PO}}\sqrt[5]{\frac{\Lambda ^{28}}{\tilde{\Lambda}^{18}}}\,\,, 
\end{align} 
where we have neglected overall numerical factors. It is then important to emphasize that the non-linear realization of boosts imposes cubic interactions of $\pi$, which lead to a bispectrum for $\zeta$. This bispectrum can be decomposed into the equilateral and orthogonal templates \cite{Senatore:2009gt} and the contribution coming from the least irrelevant operator, \textit{i.e.}\ $\smash{\dot{\pi}(\partial_i\pi)/a^2}$, has $f_{\rm NL}$ of order $\smash{\Lambda^2/(H\tilde{\Lambda})\sim (\Lambda^4/(\Delta^2_\zeta\tilde{\Lambda}^4))^{2/5}}$. 

It is also instructive to compare the parity-odd trispectrum contributions to the parity-even ones. The leading parity-even trispectrum in the Ghost Condensate is given by \citep{Izumi:2010wm} 
\be 
S_{\pi\pi\pi\pi}^{\text{(PE)}}=M_{\rm PE}^4\int {\rm d}^4x\,\sqrt{-g}\,a^{-4}(\partial_i\pi)^2(\partial_j\pi)^2\,\,.
\ee 
Let us compare this to \eqref{gi_op-1}. If we assume that all scales are of the same order, $M_{\rm PE}\sim M_{\rm PO}\sim \tilde{\Lambda} \sim {\Lambda}$, and using that $\partial_i\sim(H \Lambda)^{1/2}$ at horizon crossing due to the modified dispersion relation, we get 
\be 
\frac{\langle \pi^4\rangle_{\rm PO}}{\langle \pi^4\rangle_{\rm PE}}\Bigg|_{\text{crossing}}
\sim\frac{1}{M^4_{\rm PE}M_{\rm PO}}\frac{(H \Lambda)^{9/2}}{(H \Lambda)^{4/2}} 
\sim\Delta^2_\zeta 
\,\,, 
\ee 
meaning that in this case the parity-even trispectrum would dominate over the parity-odd one. 
Requiring that these non-Gaussianities are compatible with current bispectrum and parity-even trispectrum bounds would put a constraint on the combination $\Lambda/\tilde{\Lambda}$ (from the bispectrum) and on a combination of $\Lambda$, $\tilde{\Lambda}$ and $M_{\rm PE}$ (from the parity-even trispectrum). However, given that we are free to choose $\smash{M_{\rm PO}}$ and $\smash{\Lambda^2_{\rm PO}}$ this does not affect our conclusions about the parity-odd trispectrum. Things would be different if one could tie, via naturalness arguments, the scales $M_{\rm PE}$, $M_{\rm PO}$ and $\Lambda^2_{\rm PO}$ to $\Lambda$ and $\tilde{\Lambda}$: in this case via the constraints from the bispectrum and the (even and odd) trispectrum we could put a bound directly on all the energy scales scales in our model. We leave the exploration of naturalness in Ghost Inflation to future work.

\subsubsection{Cosmological Collider}

\noindent As discussed at the end of \S\ref{subsec:trispectrum_templates}, the size of non-Gaussianity from the diagram of \eqref{eq:exchange_diagram} depends on $c_s$. More precisely we have 
\beq
\tau_{\rm NL}^{(\lambda_1\lambda_3)} \sim 
\begin{cases}
\dfrac{c_s^4\Delta^2_\zeta\lambda_1\lambda_3}{H} &\text{for $c_s\ll1$\,\,,} \\[0.75em] 
\dfrac{\Delta^2_\zeta\lambda_1\lambda_3}{H} &\text{for $c_s=1$\,\,,} \\[0.75em]
\dfrac{\Delta^2_\zeta\lambda_1\lambda_3}{Hc_s} &\text{for $c_s\gg1$\,\,.} 
\end{cases}
\eeq 
Thanks to scale invariance this holds for any value of $\nu$. In \S\ref{sec: results} we will quote constraints on the combination $\lambda_1\lambda_3/H$ for $c_s = 0.1, 1, 10$ at different values of $\nu$. Given that the dependence on $c_s$ is particularly strong only for $c_s\ll 1$, perturbativity then requires that, at each $\nu$ value, 
\beq
    \lambda_1\lambda_3/H\lesssim10^{20}\,\,(c_s=0.1)\,\,, \qquad \text{or}\qquad\lambda_1\lambda_3/H\lesssim10^{16}\,\,(c_s\geq 1)\,\,. 
\eeq

Before concluding this section and shifting our attention to the computation of the parity-odd galaxy four-point function, we wish to comment on the fact that additional constraints on $\lambda_1$ and $\lambda_3$ come from the requirement of not having strong coupling at horizon crossing. It is straightforward to estimate this requirement by computing the EFT cutoffs associated with the interactions \eqref{CubicAction-1}\,\&\,\eqref{CubicAction-2} if $c_s=1$ and $\nu$ is close to $3/2$. These are \citep{Cabass:2022rhr}
\beq
    \smash{\Lambda_{1}\sim H^{4/3}/(\Delta^2_\zeta\lambda_{1})^{1/3}}\,\,, \qquad \smash{\Lambda_{3}\sim H/(\Delta^2_\zeta\lambda_{3})^{1/4}}\,\,,
\eeq
hence the requirement that $H/\Lambda_1\lesssim 1$, $H/\Lambda_3\lesssim 1$ is equivalent to requiring that $\tau^{(\lambda_1\lambda_3)}_{\rm NL}\Delta^2_\zeta\lesssim 1$. We expect that for $c_s\neq 1$ and generic $\nu$ the same will apply, hence we do not discuss these constraints further. 

An important constraint, instead, would come from the fact that accompanying the parity-odd contribution $\propto\lambda_1\lambda_3$ there will be two parity-even contributions $\propto\lambda_1^2$ and $\propto\lambda_3^2$. One should in principle put bounds on these three contributions simultaneously: however, we expect the contribution from gravitational non-linearities at late times will make the parity-even pieces harder to constrain. We will return to this point in \S\ref{sec: conclusion}.

\section{Galaxy Correlation Functions} 
\label{sec: data}

\noindent As demonstrated in \S\ref{sec:early_universe}, non-standard physics in inflation can source parity-violating signatures in the primordial curvature perturbation $\zeta$. As observers, however, we do not have direct access to correlators of $\zeta$, but must infer them through their late-time manifestations. For the models considered in this work, a crucial question therefore is which physical observables are sensitive to inflationary parity-violation. The usual suspects are the CMB and LSS: to capture large-scale scalar signatures like the above we require a four-point function, such as that of the CMB temperature, $\av{T^4}$, or the galaxy overdensity, $\av{\delta_g^4}$.

In this work, we probe inflationary signatures using the four-point function of spectroscopic galaxy surveys. In general, one can work either in Fourier-space (via the galaxy trispectrum) or configuration-space (via the four-point correlation function, hereafter 4PCF). From a modeling perspective, the trispectrum is preferred since there is a simple relation between the curvature perturbation and $\delta_g$; however, the 4PCF can and has been straightforwardly measured, thus it will be the focus of our attention in this work.

The galaxy 4PCF is defined as the configuration-space average of the overdensity field:
\beq
    \zeta(\vr_1,\vr_2,\vr_3) = \av{\delta_g(\vx)\delta_g(\vx+\vr_1)\delta_g(\vx+\vr_2)\delta_g(\vx+\vr_3)}_c\,\,,
\eeq
assuming homogeneity. For efficient measurement, it is useful to restrict to the isotropic component of the 4PCF (\textit{i.e.}\ that averaged over rotations), and project the statistic into a basis of spherical harmonics, defined by \citep{2020arXiv201014418C,2022MNRAS.509.2457P}:
\beq
    \zeta^{\rm iso}(\vr_1,\vr_2,\vr_3) = \sum_{\ell_1\ell_2\ell_3}\zeta_{\ell_1\ell_2\ell_3}(r_1,r_2,r_3)\P_{\ell_1\ell_2\ell_3}(\hr_1,\hr_2,\hr_3)\,\,,
\eeq
where the basis functions (related to the tripolar spherical harmonic functions introduced in \citep{1988qtam.book.....V}) are given by
\beq\label{eq: iso-basis}
    \P_{\ell_1\ell_2\ell_3}(\hr_1,\hr_2,\hr_3) = (-1)^{\ell_1+\ell_2+\ell_3}\sum_{m_1m_2m_3}\tj{\ell_1}{\ell_2}{\ell_3}{m_1}{m_2}{m_3}Y_{\ell_1 m_1}(\hr_1)Y_{\ell_2m_2}(\hr_2)Y_{\ell_3m_3}(\hr_3)\,\,,
\eeq
involving spherical harmonics and the Wigner $3j$ symbol. The 4PCF multiplets $\zeta_{\ell_1\ell_2\ell_3}(r_1,r_2,r_3)$ can be directly estimated from data (using the \textsc{encore} code \citep{2022MNRAS.509.2457P}),\footnote{\href{https://github.com/oliverphilcox/encore.git}{github.com/oliverphilcox/encore}} and are related to the full field via
\beq
    \zeta_{\ell_1\ell_2\ell_3}(r_1,r_2,r_3) = \int {\rm d}\hr_1{\rm d}\hr_2{\rm d}\hr_3\,\zeta(\vr_1,\vr_2,\vr_3)\P^*_{\ell_1\ell_2\ell_3}(\hr_1,\hr_2,\hr_3)\,\,.
\eeq
These coefficients depend on three angular momentum indices, $\ell_i$ (which satisfy triangle conditions), and three radial bins, $r_i$. Further, the basis functions are rotationally invariant, and, for even (odd) $\ell_1+\ell_2+\ell_3$ are parity-even (parity-odd). Henceforth, we will analyze the parity-odd multiplets $\zeta_{\ell_1\ell_2\ell_3}(r_1,r_2,r_3)$; in the absence of parity-violating physics in the early- or late-Universe, these are expected to be zero. 

Our primary dataset will be the observed SDSS-III BOSS DR12 galaxies \citep{2013AJ....145...10D,2011AJ....142...72E,2015ApJS..219...12A}, comprising approximately $8\times 10^5$ galaxies at redshift $z\approx 0.57$, split across the Northern and Southern galactic cap. We additionally make use of a set of $2048$ MultiDark-Patchy (hereafter \textsc{Patchy}) mock catalogs \citep{2016MNRAS.456.4156K,2016MNRAS.460.1173R}, created in order to model the noise properties of the BOSS sample, and $84$ \textsc{Nseries} mocks \citep{2017MNRAS.470.2617A}, which were introduced for BOSS pipeline validation. In all cases, we use the measured multiplets with $\ell_i\leq 4$ (satisfying the triangle conditions) and ten radial bins in $[20,160]\Mpch$, giving a total of $1288$ elements in the data-vector. The 4PCF measurements and corresponding analysis pipeline has been made publicly available on GitHub\footnote{\href{https://github.com/oliverphilcox/Parity-Odd-4PCF}{github.com/oliverphilcox/Parity-Odd-4PCF}} and further details of the dataset (including details of systematic weights and survey geometry correction) are presented in \citep{Philcox:2022hkh} (see also \citep{Hou:2022wfj}), building on the results of \citep{2021arXiv210801670P,2022MNRAS.509.2457P}.

\section{From Inflation to Galaxy Surveys: Theoretical Modeling}\label{sec: transform}

One ingredient remains in our recipe for constraining inflation with galaxy surveys; analytic predictions for the galaxy 4PCF depending on the EFTI coupling amplitudes. In this section, we will consider the relation of inflationary and late-time physics and summarize the key theoretical templates (analogously to \S{}VII of \citep{Philcox:2022hkh}). We caution that a byproduct of working in configuration-space (and in a somewhat unintuitive basis) is that the theoretical predictions are quite grotesque: as such, we relegate the finer details to Appendix \ref{app: calculations}.

\subsection{Relating $\delta_g^4$ and $\zeta^4$}
\noindent At lowest order in gravitational evolution, the galaxy density at some redshift $z$ is related to the curvature perturbation via
\beq
    \delta_g(\vk,z) = Z_1(\hk,z)M(k,z)\zeta(\vk)
\eeq
where $M(k)$ is the transfer function (including the $D_+(z)$ growth factor) and $Z_1(\hk,z)$ is the perturbative (Kaiser) kernel $b_1(z)+f(z)(\hk\cdot\hn)^2$, for linear bias $b_1$, growth rate $f$, and line of sight $\smash{\hat{\vec{n}}}$. We suppress the $z$ argument henceforth.\footnote{We neglect higher-order gravitational effects in this work. Alone, these cannot generate a parity-violating signature (without some flavor of modified gravity, e.g.~\citep{2009PhR...480....1A}), though they can serve to complicate any existing LSS templates on small scales. 
We discuss these effects in more detail in \S\ref{sec: late-time}.} The galaxy trispectrum is straightforwardly obtained in terms of the primordial trispectrum: 
\beq
    \av{\prod_{i=1}^4\delta_g(\vk_i)}_c = \left[\prod_{i=1}^4Z_1(\hk_i)M(k_i)\right]\delD{\vk_1+\vk_2+\vk_3+\vk_4}\tilde{T}(\vk_1,\vk_2,\vk_3,\vk_4)+\text{$23$ perms.}\,\,,
\eeq
using the definition \eqref{eq: Tk-def}, and, via a Fourier-transform, the (unprojected) 4PCF
\begin{equation}
\begin{split}
    \zeta(\vr_1,\vr_2,\vr_3) &= \left[\prod_{i=1}^4\int_{\vk_i}Z_1(\hk_i)M(k_i)\right]e^{i(\vk_1\cdot\vr_1+\vk_2\cdot\vr_2+\vk_3\cdot\vr_3)}\\\nonumber
    &\;\;\;\;\;\;\,\,\times\delD{\vk_1+\vk_2+\vk_3+\vk_4}\tilde{T}(\vk_1,\vk_2,\vk_3,\vk_4)+\text{23 perms.}
\end{split}
\end{equation}
Finally, we can project this onto the basis functions defined in \eqref{eq: iso-basis}, yielding
\begin{equation}
\label{eq: remapping}
\begin{split}
    \zeta_{\ell_1\ell_2\ell_3}(r_1,r_2,r_3) &= \int {\rm d}\hr_1{\rm d}\hr_2{\rm d}\hr_3\P^*_{\ell_1\ell_2\ell_3}(\hr_1,\hr_2,\hr_3)\left[\prod_{i=1}^4\int_{\vk_i}Z_1(\hk_i)M(k_i)\right]e^{i(\vk_1\cdot\vr_1+\vk_2\cdot\vr_2+\vk_3\cdot\vr_3)}\\\nonumber
    &\qquad\,\times\,\delD{\vk_1+\vk_2+\vk_3+\vk_4}\tilde{T}(\vk_1,\vk_2,\vk_3,\vk_4)+\text{23 perms.}\,\,,
\end{split}
\end{equation}
where we restrict to odd $\ell_1+\ell_2+\ell_3$ to ensure parity antisymmetry. 

In principle, \eqref{eq: remapping} contains all the details needed to compute a theoretical template for the observed galaxy 4PCF using the inflationary correlators of \S\ref{sec:early_universe}. In practice, this is highly non-trivial, due to the $18$-dimensional coupled integrals. A number of tricks can be used to simplify this, as detailed in Appendix \ref{app-sub: tricks}. In brief: (a) the $\hr$ integrals can be performed analytically using spherical harmonic orthogonality, (b) we can rewrite the Dirac delta function as a one- or two-dimensional integral, and (c) $Z_1(\hk)$ can be expressed as a spherical harmonic series. Computation then reduces to a set of radial integrals with an associated angular piece depending only on spherical harmonics in $\hk$, which can be expressed in terms of Wigner $3j$ and $9j$ symbols, aided by writing the expanding the various terms in a rotationally-invariant basis. All in all, we will arrive at an expression for the relevant template written in terms only of low-dimensional integrals and angular momentum couplings. We summarize the corresponding templates for the Ghost Inflation and Cosmological Collider models below.

\subsection{Ghost Inflation}\label{subsec: GI-template}
\noindent Inserting the templates of \eqref{eq: primordial-ghost} into the late-time definition \eqref{eq: remapping}, we can obtain the 4PCF templates for the two Ghost Inflation correlators considered in \S\ref{sec:early_universe}. Following the simplifications outlined in Appendix \ref{app-sub: GI}, these can be written
\beq\label{eq: ghost-4pcf-I}
	\zeta^{(M_{\rm PO})}_{\ell_1\ell_2\ell_3}(r_1,r_2,r_3) &=& 
	2(4\pi)^{11/2}(-i)^{\ell_{123}}\frac{\Lambda^5(H\tilde\Lambda)^{1/2}}{M_{\rm PO}\tilde\Lambda^5\Gamma(\tfrac{3}{4})^2}(\Delta_\zeta^2)^3\sum_H\Phi_H\sum_{L_1\cdots L_4L'}(-i)^{L_{1234}}\tjo{L_1}{L_2}{L'}\tjo{L'}{L_3}{L_4}\nonumber\\
	&&\,\times\,\mathcal{C}_{L_1L_2L_3L_4L'}\mathcal{M}_{L_1L_2(L')L_3L_4}^{\ell_{H1}\ell_{H2}(\ell')\ell_{H3}\ell_{H4}}\\\nonumber
	&&\,\times\,\mathrm{Im}\int_0^\infty x^2 {\rm d}x\int_0^\infty\lambda^{11}{\rm d}\lambda\,I_{3/4,1/2,\ell_{H1},L_1}(x,\lambda;r_{H1})I_{3/4,3/2,\ell_{H2},L_2}(x,\lambda;r_{H2})\\\nonumber
	&&\qquad\qquad\qquad\qquad\times\,I_{3/4,1/2,\ell_{H3},L_3}(x,\lambda;r_{H3})I_{3/4,1/2,\ell_{H4},L_4}(x,\lambda;r_{H4})\,\,,
\eeq
and
\beq\label{eq: ghost-4pcf-II}
	\zeta^{(\Lambda_{\rm PO}^2)}_{\ell_1\ell_2\ell_3}(r_1,r_2,r_3) &=& 
	\frac{8\sqrt{2}}{3\sqrt{5}}(4\pi)^{11/2}(-i)^{\ell_{123}}\frac{\Lambda^5(H\tilde\Lambda)^{3/2}}{\Lambda^2_{\rm PO}\tilde\Lambda^6\Gamma(\tfrac{3}{4})^2}(\Delta_\zeta^2)^3\sum_H\Phi_H\sum_{L_1\cdots L_4L'}(-i)^{L_{1234}}\tjo{L_1}{L_2}{L'}\tjo{L'}{L_3}{L_4}\nonumber\\
	&&\,\times\,\mathcal{C}_{L_1L_2L_3L_4L'}\mathcal{N}_{L_1L_2(L')L_3L_4}^{\ell_{H1}\ell_{H2}(\ell')\ell_{H3}\ell_{H4}}\\\nonumber
	&&\,\times\,\int_0^\infty x^2 {\rm d}x\int_0^\infty\lambda^{13}{\rm d}\lambda\,I_{3/4,1/2,\ell_{H1},L_1}(x,\lambda;r_{H1})I_{3/4,5/2,\ell_{H2},L_2}(x,\lambda;r_{H2})\\\nonumber
	&&\qquad\qquad\qquad\qquad\times\,I_{3/4,3/2,\ell_{H3},L_3}(x,\lambda;r_{H3})I_{-1/4,1/2,\ell_{H4},L_4}(x,\lambda;r_{H4})\,\,,
\eeq
involving the $I$ integrals (over the transfer and Hankel functions, see \ref{eq: ghostI}), $\mathcal{C}_{L_1\cdots L_n}\equiv \sqrt{(2L_1+1)\cdots(2L_n+1)}$, a permutation factor $\Phi_H\in\{\pm1\}$, and the coupling matrices $\mathcal{M}$ and $\mathcal{N}$ of \eqref{eq: ghost-coupling}, which can be expressed in terms of Wigner $9j$ symbols.

To facilitate efficient computation of the above templates, we employ a number of tricks. Firstly, for the $k$ integrals, we can assess their convergence via the asymptotic limits of the component functions, \textit{i.e.}\ 
\beq
	j_\ell(x) \sim \frac{\sin(x-\ell\pi/2)}{x}\,\,, \qquad H_\alpha^{(1)}(2ix^2) \sim \frac{(-i)^{1+\alpha}}{\sqrt{\pi}x}e^{-2x^2} 
\eeq
for $x\gg1$. The integrand of the $I$ functions thus behaves as
\beq
	\frac{k^{2+\beta}}{2\pi^2}M(k)j_\ell(kr)j_L(kx)H^{(1)}_{\alpha}(2ik^2\lambda^2)\sim (-i)^{1+\alpha}\frac{k^{\beta-1}}{2\pi^{5/2}}\frac{M(k)}{xr\lambda}\sin\left(kr+\ell\pi/2\right)\sin\left(kx+L\pi/2\right)e^{-2k^2\lambda^2}\,\,,
\eeq
for $kx,kr\gg\ell$, which is exponentially convergent. In this vein, it is useful change variables in the $k$ integral to $q\equiv k\lambda$: thence, from \eqref{eq: ghostI},
\beq
	I_{\alpha,\beta,\ell,L}(x,\lambda;r) = \lambda^{-3-\beta}\int_0^\infty\frac{q^{2+\beta}{\rm d}q}{2\pi^2}M(q/\lambda)j_\ell(qr/\lambda)j_L(qy)H^{(1)}_{\alpha}(2iq^2)\,\,,
\eeq
defining also $y=x/\lambda$, which ensures that an appropriate range of $x$ values can be used for any $\lambda$. Additionally, we note that the $y$-integrals can be rewritten as (infinite) discrete summations, if one imposes some maximum $q$ of interest, \textit{i.e.}\, 
\beq
	\int_0^\infty y^2{\rm d}y\,\prod_{j=1}^4j_{L_j}(q_jy) = \left(\frac{\pi}{2q_{\rm max}}\right)^3\sum_{m=0}^\infty m^2 \mathcal{E}_m\prod_{j=1}^4j_{\ell_j}\left(\frac{q_j\pi m}{2q_{\rm max}}\right)\,\,,
\eeq
where $\mathcal{E}_m$ is $1/2$ if $m=0$ and unity else \citep{2021RSPSA.47710376P}. In practice, this was not found to significantly expedite computation.

When evaluating the 4PCF, we additionally integrate the radial components over finite bins, matching that of the data. This is achieved via the replacement:
\beq
	j_\ell(kr) \to \frac{1}{V_{\rm bin}}\int_{r_{\rm min}}^{r_{\rm max}}r^2dr\,j_\ell(kr)\,\,,
\eeq
where the radial bin is specified by $[r_{\rm min},r_{\rm max}]$ with volume $V_{\rm bin}$. The bin-integrated Bessel functions are analytic and can be found in \citep{2020MNRAS.492.1214P}. 

To evaluate the theoretical model, we assume an integration grid of $200$, $500$ and $500$ points in $q$, $y$, and $\lambda$ respectively, with maximum values of $5$, $100$ and $250$, verified by initial testing. To match the data, we use $10$ radial bins in $r$, linearly spaced in $[20,160]\Mpch$, and all odd multiplets up to $\ell_{\rm max}=4$, with $\ell_{\rm max}=8$ used for all internal ($L_i$) summations. The full computation required $\approx 48$ hours on a 24-core machine.

\subsection{Cosmological Collider}\label{subsec: CC-template}
\noindent The particle-exchange 4PCF can be obtained in a similar manner, inserting the trispectrum definition \eqref{eq: primordial-collider-oliver} into \eqref{eq: remapping} and simplifying. The full calculation, outlined in Appendix \ref{app-sub: CC}, yields
\beq\label{eq: CC-4pcf}
    \zeta^{(\lambda_1\lambda_3)}_{\ell_1\ell_2\ell_3}(r_1,r_2,r_3) &=& (4\pi)^{7/2}(-i)^{\ell_{123}}\frac{c_s^4\lambda_1\lambda_3}{\resub{18}\sqrt{5}H}(\Delta_\zeta^2)^4\sin\pi\left(\nu+\frac{1}{2}\right)\sum_H\Phi_H\sum_{L_1\ldots L_4L'}(-i)^{L_{1234}}\tjo{L_1}{L_2}{L'}\tjo{L'}{L_3}{L_4}
    \nonumber\\
    &&\,\times\,\mathcal{C}_{L_1\ldots L_4L'}\,\mathcal{O}_{L_1L_2(L')L_3L_4}^{\ell_{H1}\ell_{H2}(\ell')\ell_{H3}\ell_{H4}}\int\frac{s^2{\rm d}s}{2\pi^2}Q^{\ell_{H1}\ell_{H2},A}_{L_1L_2L'}(s;r_{H1},r_{H2})Q^{\ell_{H3}\ell_{H4},B}_{L_3L_4L'}(s;r_{H3},r_{H4})\,\,,
\eeq
where the $Q$ integrals are defined in \eqref{eq: colliderQ}, and the coupling matrix, $\mathcal{O}$, is given in \eqref{eq: couplingO}.

Na\"ive computation of \S\ref{subsec: GI-template} is difficult since the $Q$ functions involve integrals of three sets of spherical Bessel functions, which are highly oscillatory, and each must be then integrated over two momentum variables. These can be simplified using the relation of \citep{1991JPhA...24.1435M}:
\beq
    \tjo{L_1}{L_2}{L'}\int x^2{\rm d}x\,j_{L_1}(k_1x)j_{L_2}(k_2x)j_{L'}(sx) &=& \frac{\pi\beta(\Delta)}{4k_1k_2s}i^{L_1+L_2-L'}(2L'+1)^{1/2}\left(\frac{k_1}{s}\right)^{L'}\sum_{\lambda=0}^{L'}\binom{2L'}{2\lambda}^{1/2}\left(\frac{k_2}{k_1}\right)^{\lambda}\\\nonumber
    &&\,\times\,\sum_\ell (2\ell+1)\tjo{L_1}{L'-\lambda}{\ell}\tjo{L_2}{\lambda}{\ell}\begin{Bmatrix}L_1 & L_2 & L'\\ \lambda & L'-\lambda & \ell\end{Bmatrix}\mathcal{L}_\ell(\Delta)\,\,,
\eeq
where the curly braces indicate Wigner $6j$ symbols, $\Delta = (k_1^2+k_2^2-s^2)/(2k_1k_2)$, and $\beta(\Delta)$ is unity if $|\Delta|< 1$ and zero else. This reduces the $Q$ integrals to the form:
\beq
    Q^{\ell_1\ell_2,X}_{L_1L_2L'}(s;r_1,r_2) &=& f_{L_1L_2L'}(k_1,k_2,s)\left[\int\frac{k_1^2{\rm d}k_1}{2\pi^2}M(k_1)\bar{j}_{\ell_{1}}(k_1;r_1)\right]\left[\int\frac{k_2^2{\rm d}k_2}{2\pi^2}M(k_2)\bar{j}_{\ell_{2}}(k_2;r_2)\right]t^{X}(k_1,k_2,s),
\eeq
where $f_{L_1L_2L'}(k_1,k_2,s)$ is the result of the above Bessel function integral, and we have additionally replaced the Bessel functions in $k_ir_i$ by their bin-integrated forms, as in \S\ref{subsec: GI-template}. This may be computed numerically as a two-dimensional integral in $(k_1,k_2)$ for a grid of values of $s$ and all radial and angular bins of interest. In practice, we utilize an integration grid of $100$ points in $k_i$ and $100$ points in $s$ with maximal values of $k_{\rm max} = 0.5\hMpc$ and $s_{\rm max}=1.0\hMpc$ (noting that the Bessel functions have support for $k\lesssim 1/R_{\rm min}$ and $s\leq 2k_{\rm max}$ from the triangle condition). Computation requires $\approx 36$ hours on a $24$-core machine for each choice of sound-speed and mass parameter.

\section{From Galaxy Surveys to Inflation: Results}
\label{sec: results}

\noindent Our primary goal in this work is to use the measured parity-odd four-point correlation function to search for signatures of new inflationary physics, such as massive particle exchange. In brief, our approach is to constrain the amplitude, $A$, of a given model by comparing the measured and theoretical four-point functions (denoted $A\zeta_{\rm th}$ and $\zeta_{\rm obs}$) via the following likelihood:
\beq\label{eq: lik-gauss}
    L(A) \propto \mathrm{exp}\left(-\frac{1}{2}\left[A\mathbb{P}\zeta_{\rm th} - \mathbb{P}\zeta_{\rm obs}\right]^TC^{-1}\left[A\mathbb{P}\zeta_{\rm th} - \mathbb{P}\zeta_{\rm obs}\right]\right)\,\,,
\eeq
where $\mathbb{P}$ is a projection matrix used to reduce the dimensionality of the 4PCF, and $C = \av{(\mathbb{P}\zeta)(\mathbb{P}\zeta)^T}$ is the covariance, measured from simulations. Via posterior sampling, we can compute the constraints on $A$, and thus evaluate the viability of a given theoretical model.

In more detail, we project the 4PCF onto a low-dimensional basis defined by first computing the eigendecomposition of the theoretical covariance matrix \citep{2021arXiv210801714H}, then selecting the basis vectors which maximize the signal-to-noise of the theoretical model $\zeta_{\rm th}$. In general, we will assume $N_{\rm eig}=100$ basis vectors, following \citep{Philcox:2022hkh}, in order to avoid any potential loss-of-information (at low $N_{\rm eig}$) with non-Gaussianity of the likelihood (at high $N_{\rm eig}$), due to highly correlated data.

For the covariance, we utilize measurements of the 4PCF from the $N_{\rm mocks}=2048$ \textsc{Patchy} mocks, defined as
\beq
    C = \frac{1}{N_{\rm mocks}-1}\sum_{a=1}^{N_{\rm mocks}}\left(\mathbb{P}\zeta^{(a)}\right)\left(\mathbb{P}\zeta^{(a)}\right)^T\,\,,
\eeq
where $(a)$ indicates the measurement from mock $a$. To account for the presence of noise in the covariance matrix, we use the following likelihood, instead of the Gaussian \eqref{eq: lik-gauss} (which applies in the $N_{\rm mocks}\to\infty$ limit):
\beq\label{eq: lik-SH}
    L(A) \propto \left(1+\frac{\left[A\mathbb{P}\zeta_{\rm th} - \mathbb{P}\zeta_{\rm obs}\right]^TC^{-1}\left[A\mathbb{P}\zeta_{\rm th} - \mathbb{P}\zeta_{\rm obs}\right]}{N_{\rm mocks}-1}\right)^{-N_{\rm mocks}/2}\,\,,
\eeq
as discussed in \citep{2016MNRAS.456L.132S}. All posterior sampling is performed using \textsc{emcee} \citep{Foreman-Mackey:2012any} in the one-dimensional parameter space.

\subsection{Ghost Inflation}

\noindent For the Ghost Inflation scenario, we constrain the following amplitudes, extracted from the prefactors of \eqref{eq: primordial-ghost}:
\beq
    A^{(M_{\rm PO})}\equiv\frac{\Lambda^5H^{1/2}}{M_{\rm PO}\tilde\Lambda^{9/2}}\times\frac{(\Delta^2_{\zeta})^3}{\Gamma(\tfrac{3}{4})^2}\,\,, \qquad A^{(\Lambda^2_{\rm PO})}\equiv\frac{\Lambda^5H^{3/2}}{\Lambda^2_{\rm PO}\tilde\Lambda^{9/2}}\times\frac{(\Delta^2_{\zeta})^3}{\Gamma(\tfrac{3}{4})^2}\,\,,
\eeq
where we separate out terms appearing in the $\tau_{\rm NL}$ definitions of \eqref{tauNL_in_Hubble}.

\begin{figure}
    \centering
    \includegraphics[width=0.9\textwidth]{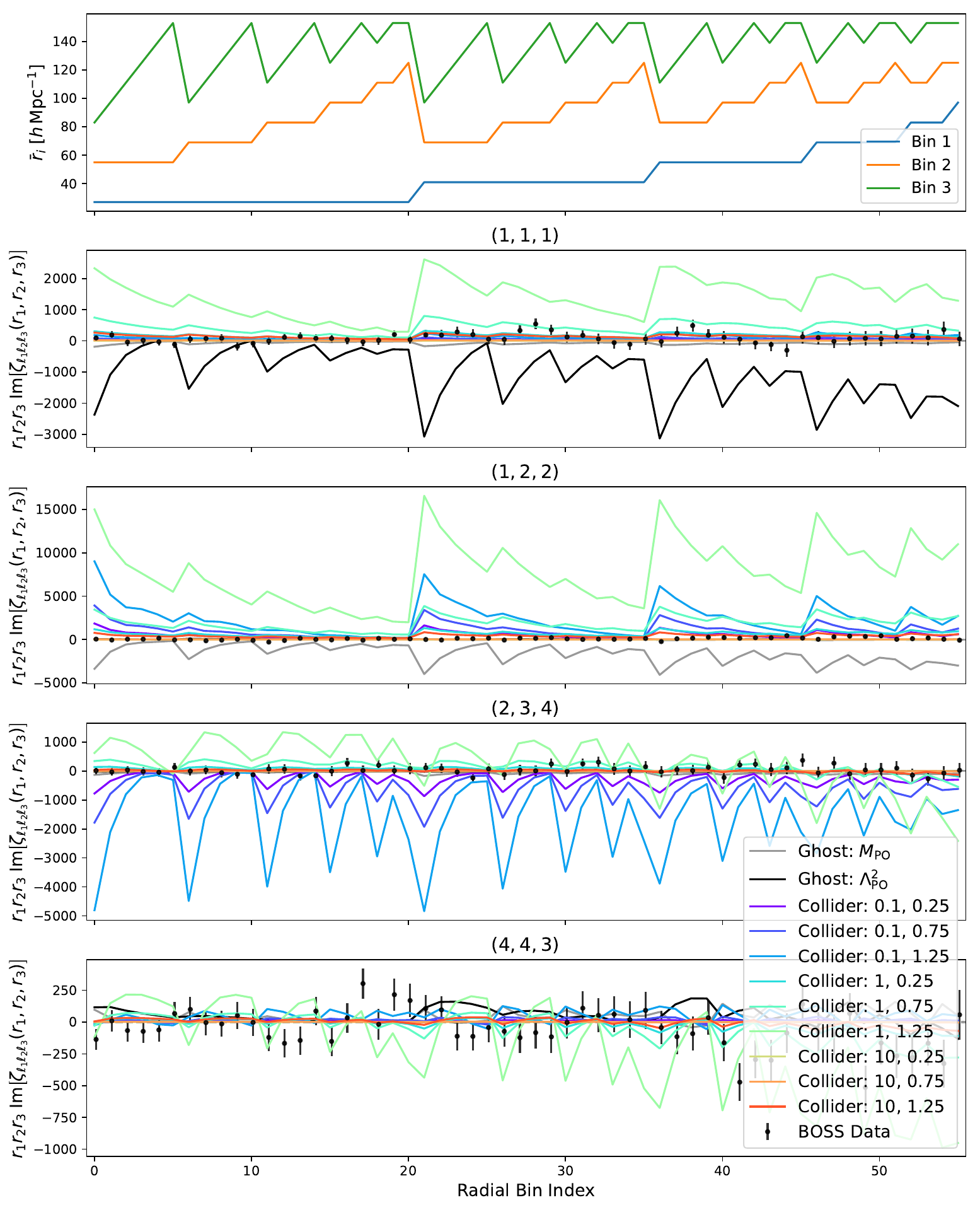}
    \caption{Comparison of the observed galaxy four-point correlation function, $\zeta_{\ell_1\ell_2\ell_3}(r_1,r_2,r_3)$ (black points), with two models of the theoretical 4PCF, assuming Ghost Condensate (black lines) and Cosmological Collider (colored lines) inflation. We assume fiducial values of $A^{(M_{\rm PO})}=100A^{(\Lambda^2_{\rm PO})}=10^{-10}$ and $A^{(\lambda_1\lambda_3)} = 10^{19}(\Delta_\zeta^2)^4c_s^4\sin\pi\left(\nu+\frac{1}{2}\right)$ for visibility and consider a variety of values of the sound speed $c_s$ and mass $\nu$ in the latter case, indicated by the captions. The second through fifth panels show the correlators for a selection of values of $\ell_1,\ell_2,\ell_3$ (indicated by the title), with the $x$-axis giving the radial bins, collapsed into one dimension. The first panel shows the values of the radial bin centers corresponding to each one-dimensional bin center. Here, we utilize data from the BOSS CMASS NGC region, and with error-bars obtained from the \textsc{Patchy} simulations (noting that the data is highly correlated). Notably, the theoretical models have strong (and different) dependence on the multiplet, which the MCMC analysis shows to be broadly inconsistent with the data. Constraints on the model amplitudes are given in Tab.\,\ref{tab: amplitude-constraints}\,\&\,\ref{tab: collider-amplitudes}.}
    \label{fig: ghost-4pcf}
\end{figure}

In Fig.~\ref{fig: ghost-4pcf}, we plot a comparison of the observed galaxy 4PCF (from \citep{Philcox:2022hkh}) and the 4PCF Ghost Inflation prediction, utilizing values of $A^{(M_{\rm PO})} = 100 A^{(\Lambda_{\rm PO}^2)}=10^{-10}$ for visibility. Notably, the theoretical models have strong dependence on both the radial and angular parameters ($r_i$ and $\ell_i$), and the fiducial values are clearly inconsistent with the data. Enhanced signals are seen particularly for small multiplets (corresponding to wide angles) and the smaller radial bins, though we see features also at high $r$.

\begin{figure}
    \centering
    \includegraphics[width=0.47\textwidth]{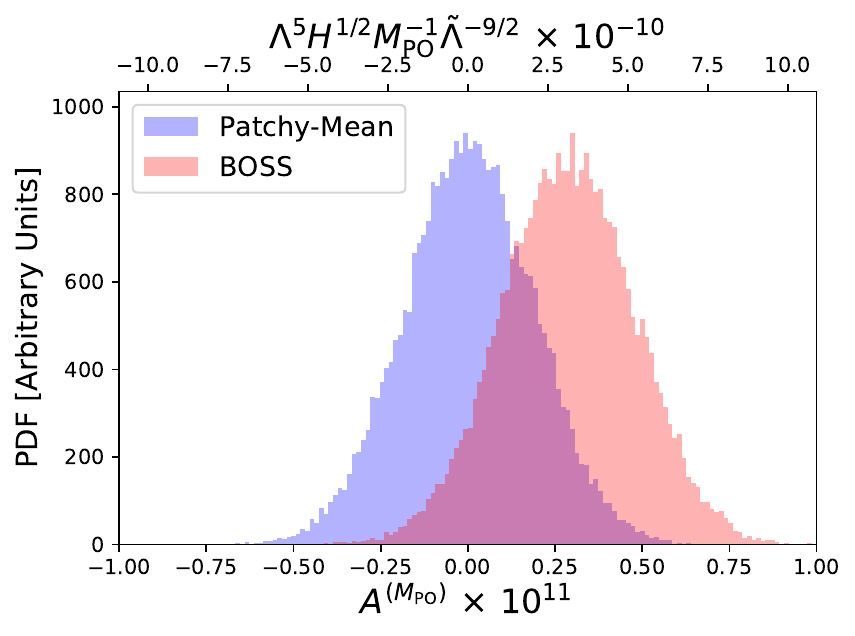}
    \includegraphics[width=0.47\textwidth]{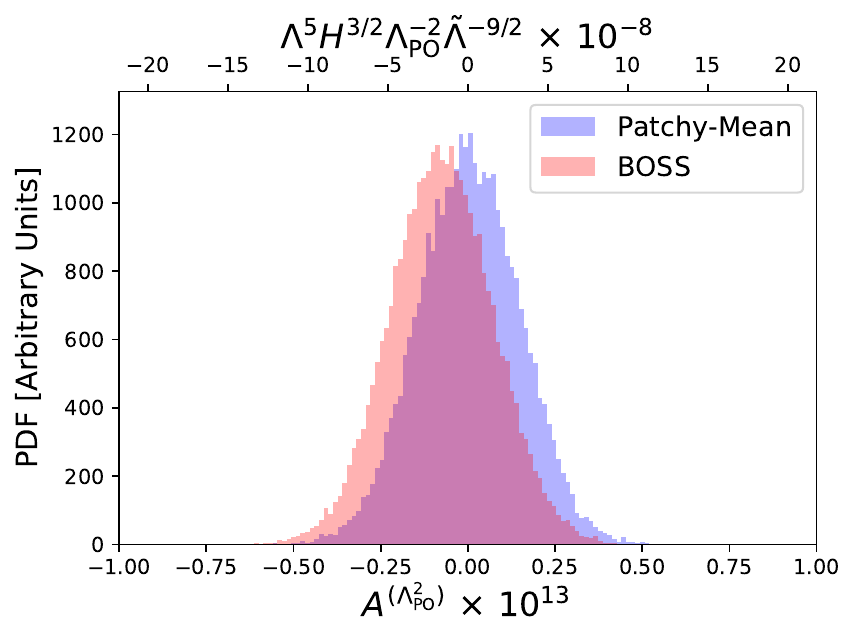}
    \caption{Constraints on the amplitude of parity-violation in the Ghost Condensate inflationary model, from the leading (left) and subleading (right) EFTI diagrams. We show results both from the mean of 2048 \textsc{Patchy} mocks (blue) and the BOSS data (red), constraining both $A^{(M_{\rm PO})}, A^{(\Lambda^2_{\rm PO})}$ and the corresponding physical parameters, as shown in the title. The $1\sigma$ constraints are shown in Tab.\,\ref{tab: amplitude-constraints}, and correspond to the combinations $\Lambda^5H^{1/2}M_{\rm PO}^{-1}\tilde{\Lambda}^{-9/2}=(3.1\pm2.1)\times 10^{10}$ and $\Lambda^5H^{3/2}\Lambda_{\rm PO}^{-2}\tilde{\Lambda}^{-9/2}=(-1.5\pm8.1)\times 10^8$.}
    \label{fig: ghost-amplitude}
\end{figure}

To explore the feasibility of the general Ghost Inflation models, we perform an MCMC analysis to find constraints on $A^{(M_{\rm PO})}, A^{(\Lambda_{\rm PO}^2)}$, as described above. Since the two operators arise at different orders in the EFTI, we will consider the templates separately rather than performing a joint analysis. The results are shown in Fig.\,\ref{fig: ghost-amplitude} and Tab.\,\ref{tab: amplitude-constraints}. Analyzing the mean 4PCF of the \textsc{Patchy} and \textsc{Nseries} mock catalogs, we find a ghost amplitude highly consistent with zero; this is a good consistency test of our analysis, particularly since the \textsc{Nseries} catalogs are high-resolution mocks and include various physical effects such as redshift-space distortions, survey windows, and fiber collision artefacts. For the BOSS data, we find the $68\%$ confidence intervals $A^{(M_{\rm PO})}=(1.4\pm1.0)\times 10^{-12}$ and $A^{(\Lambda_{\rm PO}^2)}=(-0.7\pm3.7)\times 10^{-14}$, or the physical constraints $\Lambda^5H^{1/2}M_{\rm PO}^{-1}\tilde{\Lambda}^{-9/2}=(3.1\pm2.1)\times 10^{10}$ and $\Lambda^5H^{3/2}\Lambda_{\rm PO}^{-2}\tilde{\Lambda}^{-9/2}=(-1.5\pm8.1)\times 10^8$ assuming $\Delta_\zeta^2=4.1\times 10^{-8}$ \citep{2020A&A...641A...6P}, giving no evidence for a parity-violating Ghost Condensate. 

Finally, we can compare the above results to the perturbativity bounds discussed in \S\ref{subsec: perturbativity}, through the limits on $\tau_{\rm NL}^{(M_{\rm PO})}$, and $\tau^{(\Lambda^2_{\rm PO})}_{\rm NL}$, which correspond to $\Lambda^5H^{1/2}M_{\rm PO}^{-1}\tilde\Lambda^{-9/2}$ and $\Lambda^5H^{3/2}\Lambda_{\rm PO}^{-2}\tilde\Lambda^{-9/2}$, up to numerical constants. Perturbativity requires $\tau_{\rm NL} \lesssim 10^{8}$, which is of the same order as the constraints above. Precise statements involving these bounds are difficult, given that they are derived only up to numerical factors; however, this indicates that we are working in the weakly non-Gaussian regime, where the EFTI expansion is valid. Furthermore, this implies that the BOSS constraints on the Lagrangian amplitudes are parametrically relevant with regards to the values predicted within the EFTI.

\begin{table}[]
    \centering
    \begin{tabular}{l|c|c||c}
     \textbf{Dataset} & \quad $10^{12}A^{(M_{\rm PO})}$ \quad & \quad $10^{14}A^{(\Lambda^2_{\rm PO})}$ \quad & 
     \quad \resub{$10^{13}A^{(\lambda_1\lambda_3)}(1,1)$} \quad  \\\hline\hline
     BOSS     &  $1.4\pm1.0$ & $-0.8\pm3.8$ & 
     \resub{$-3.1\pm4.1$}\\
     \textsc{Patchy} & $0.0\pm0.9$ & $0.0\pm3.9$ & 
     $\resub{0.1\pm 4.0}$\\
     \textsc{Nseries} & $0.2\pm1.1$ & $0.3\pm4.7$ & 
     \resub{$0.4\pm4.4$}\\
    \end{tabular}
    \caption{Constraints on the amplitudes of parity-violating inflationary models using the observational 4PCF data. The left panels show results for Ghost Inflation, whilst the right give those for the Cosmological Collider using a single value of particle mass and sound speed. In all cases, we quote $68\%$ confidence intervals.
    For the collider model, we assume the parameters $\nu = c_s = 1$; constraints on the amplitude for a range of parameter values are given in Tab.\,\ref{tab: collider-amplitudes}. We give results from BOSS, the mean of $2048$ \textsc{Patchy} mocks, and the mean of $84$ \textsc{Nseries} mocks, with the former two being shown in Fig.\,\ref{fig: ghost-amplitude} (for the Ghost Condensate) and Fig.\,\ref{fig: collider-amplitude} (for the Cosmological Collider). We find no detection of any parity-violating model.}
    \label{tab: amplitude-constraints}
\end{table}

\subsection{Cosmological Collider}

\noindent When analyzing the Cosmological Collider 4PCF, we instead constrain the amplitude
\beq
    A^{(\lambda_1\lambda_3)}(c_s,\nu) \equiv \frac{\lambda_1\lambda_3}{H}\times c_s^4(\Delta_\zeta^2)^4\sin\pi\left(\nu+\frac{1}{2}\right)
\eeq
given some values of $c_s$ and $\nu$, again separating out the part appearing in $\tau_{\rm NL}$. As noted in \S\ref{sec:early_universe}, this vanishes for massless and conformally coupled particles (at $\nu = 3/2, 1/2$) which do not yield a parity-violating signature. In Fig.\,\ref{fig: ghost-4pcf}, we plot the collider model alongside the BOSS data with a fiducial value $\lambda_1\lambda_3/H=1\times 10^{19}$. Interestingly, the models exhibit significantly different scale-dependence to Ghost Inflation, with a particular enhancement seen in the higher multiplets with respect to $\zeta_{111}$, though we once again see enhanced signals on large scales.\footnote{It would be interesting, though beyond the scope of this work, to study how much this difference is due to the fact that the collider trispectrum arises from an exchange diagram, rather than a contact diagram.} Furthermore, the templates vary considerably with $c_s$, both in sign, amplitude, and scale-dependence. As predicted (see the discussion in \S\ref{subsec: perturbativity}), the signatures are largest when both fields have the same sound speed ($c_s = 1$), indicating that the constraints will be tightest on such models. Purely from visual inspection, none appear to be consistent with the data. 

More rigorously, we may again perform parameter inference to constrain the amplitude of the parity-breaking Cosmological Collider coupling. In this case, the model strictly depends on three parameters: the coupling strength, $\lambda_1\lambda_3/H$, the sound speed of the heavy particle, $c_s$, and the mass, parametrized by $\nu$. Rather than scan over all three, we here constrain only the amplitude for a variety of fixed values of $c_s$ and $\nu$, noting that the latter parameters do not enter the model linearly, thus are difficult to scan over efficiently. 

\begin{figure}
    \centering
    \includegraphics[width=0.6\textwidth]{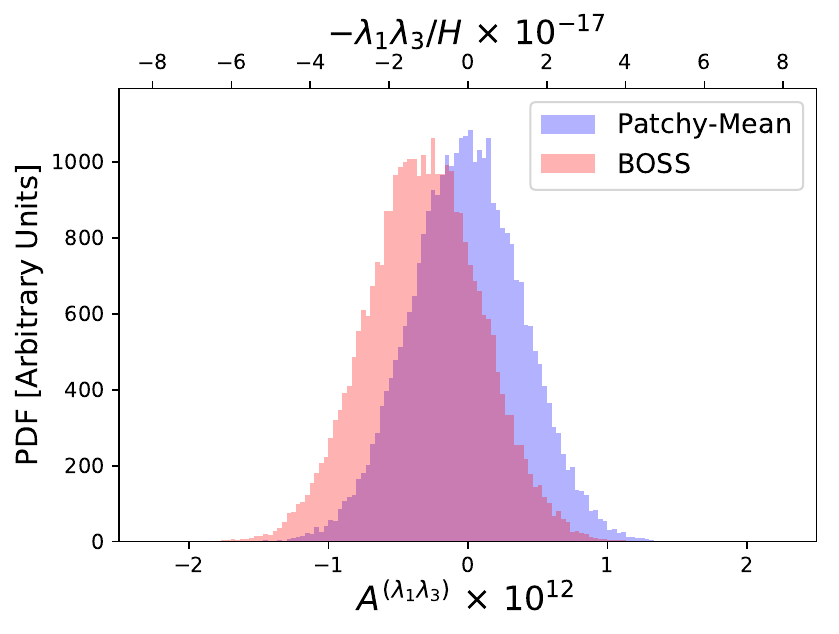}
    \caption{As Fig.\,\ref{fig: ghost-amplitude}, but constraining the amplitude of parity-violation in the Cosmological Collider inflationary model, for a massive spin-$1$ particle. Here, we use parameters $c_s = 1$ and $\nu = 1$, \textit{i.e.}\ $m_\sigma = \sqrt{5}H/2$. The $1\sigma$ constraints are shown in Tab.\,\ref{tab: amplitude-constraints}, and correspond to the combination $\lambda_{1}\lambda_{3}/H=(\resub{1.1\pm1.4})\times 10^{17}$.}
    \label{fig: collider-amplitude}
\end{figure}

First, we consider the constraints from a single model with $\nu=c_s=1$ (corresponding to a $m_\sigma=\sqrt{5}H/2$ spin-$1$ field with the same sound-speed as the inflaton). The resulting bounds on $A^{(\lambda_1\lambda_3)}(c_s,\nu)$ are shown in Fig.\,\ref{fig: collider-amplitude} and Tab.\,\ref{tab: amplitude-constraints}. As for Ghost Inflation, the amplitude is consistent with zero for the BOSS data. Moreover, inference on the mean of \textsc{Patchy} and \textsc{Nseries} simulations also returns a null result, implying that a spurious inflationary signal is not generated in our modeling, analysis pipeline, or systematic treatment at any detectable level. In this case, the BOSS data constraints the coupling $\lambda_{1}\lambda_{3}/H =(\resub{1.1\pm1.4})\times 10^{17}$ at $68\%$ confidence.

Tab.\,\ref{tab: collider-amplitudes} gives the analogous constraints on $\lambda_1\lambda_3/H$ for a range of values of $\nu$ and $c_s$ consistent with the physical bounds. Whilst we display results only for BOSS, we have repeated the analysis also for the two simulation suites and find null detections in all cases. Constraints on the coupling strength vary both as a function of $c_s$ and $\nu$: as expected, we observe somewhat stronger constraints for $c_s=1$ and the bounds tighten slightly as $\nu$ increases (or the mass decreases). Of course, for massless and conformally coupled particles, our bounds are infinite since no parity-violating inflationary trispectrum is generated. In all cases, our constraints are below $2\sigma$, indicating no significant evidence for any model. Furthermore, the significance of any detection is reduced due to the look-elsewhere effect due to analyzing a large number of models \citep[e.g.][]{Bayer:2020pva}, and we note that each analysis is far from independent. Overall, we conclude that the inflationary exchange of a massive spin-$1$ particle does not seem to source parity-violation in the galaxy correlator at any currently-detectable level.

As before, these results may be compared to the perturbativity constraints discussed in \S\ref{subsec: perturbativity}. In general, this, and the restriction that there is no strong coupling at horizon crossing, demands that $\lambda_1\lambda_3/H\lesssim 10^{20}$ (for $c_s=0.1$) or $10^{16}$ (for $c_s\geq 1$). The values reported in Tab.\,\ref{tab: collider-amplitudes} are roughly consistent with this (noting that we have neglected powers of $\pi^4$, et cetera), particularly for $c_s=1$. As for Ghost Inflation, this implies that our constraints are consistent with the EFTI framework, and that constraints from future surveys are expected to be phenomenologically relevant.

\begin{table}[!htb]
    \centering
    \begin{tabular}{c||c|c|c|c|c}
    & $\nu=0$ & $\nu=0.25$ & $\nu=0.75$ & $\nu=1$ & $\nu=1.25$ \\\hline\hline
    $c_s = 0.1$ & 
    $(-5 \pm 7) \times 10^{17}$ & 
    $(-0.7 \pm 1) \times 10^{18}$ &
    $(3 \pm 8) \times 10^{17}$ &
    $(6 \pm 5) \times 10^{17}$ &
    $(5 \pm 5) \times 10^{17}$ \\
    $c_s = 1$ &
    $(0.8 \pm 5) \times 10^{17}$ 
    & $(1 \pm 7) \times 10^{17}$ 
    & $(0.8 \pm 4) \times 10^{17}$ 
    & $(0.2 \pm 2) \times 10^{17}$ 
    & $(-0.7 \pm 2) \times 10^{17}$ \\
    $c_s = 10$ & 
    $(-7 \pm 9) \times 10^{19}$ &
    $(-1 \pm 1) \times 10^{20}$ &
    $(-0.3 \pm 3) \times 10^{19}$ &
    $(-0.8 \pm 7) \times 10^{18}$ &
    $(0.5 \pm 2) \times 10^{18}$ 
    \end{tabular}
    \caption{Constraints on the coupling strength $\lambda_{1}\lambda_{3}/H$ of spin-$1$ massive particles inflation using the parity-violating 4PCF from BOSS galaxies. We give results for a variety of values of $\nu\equiv \sqrt{9/4-(m_\sigma/H)^2}$ and the massive particle sound-speed, $c_s$. Here, results are given only for the BOSS data; a representative result for the \textsc{Patchy} and \textsc{Nseries} simulations is shown in Tab.\,\ref{tab: amplitude-constraints}. We omit $\nu = 1/2, 3/2$ which do not give parity-violating signatures. We find no significant detection of parity-violation in any case (with all values under $1.4\sigma$), though note that the various templates are highly correlated.}
    \label{tab: collider-amplitudes}
\end{table}


\section{Late-Time Parity-Violation and the EFT of Large Scale Structure} 
\label{sec: late-time}

\noindent Until now, we have considered only early-Universe sources of parity-violation. One may also ask the following question: how large can parity-violating signatures be in the late Universe? The EFTofLSS, which allows for the description of structure formation on large scales in terms of a weakly-coupled theory even when the details of complicated baryonic physics governing galaxy formation are unknown, naturally provides an answer to this question. Below, we will examine the various ways late-time parity-violation could enter the picture within this framework. One caveat should be highlighted: the terms below require a parity-violating mechanism for their production (such as some flavor of chiral gravity). In the absence of this, the bias coefficients accompanying each parity-violating term will be exactly zero, thus there will be no late-time effects.

\subsection{Density Contributions}

\noindent Within the EFTofLSS, the galaxy overdensity, $\delta_g$, is represented in terms of operators built out of the matter density, tidal fields, and their derivatives. \resub{Assuming statistical homogeneity, statistical isotropy, and the equivalence principle}, we can construct a complete basis at any order in perturbations. This expansion is perturbative and controlled by several different length scales: in the galaxy rest frame, these are $k_{\rm NL}$, the scale at which the gravitational collapse of matter becomes fully non-linear, and $R_\ast$, which depends both on the details of the host halo formation and baryonic physics that affects galaxy formation. After factoring out the relevant length scales, the bias expansion depends on a number of free coefficients (bias parameters), which have to be determined experimentally, \textit{i.e.}\ fitted from data.

Let us discuss how  parity-violating four-point function may arise in the EFTofLSS. We first work in the galaxy rest frame, whereupon the bias expansion takes the form 
\beq
\delta_g(\vec{x}) = \sum_{n=1}^{}\sum_{\cal O}b^{(n)}_{\cal O}{\cal O}^{(n)}(\vec{x})\,\,,
\eeq
where $\{\mathcal{O}\}$ are various bias operators and the index $n$ means that a given operator starts at order $n$ in the (non-linear) matter density field. Non-linearities in the bias expansion can generate non-trivial four-point functions, even if none are present in the primordial density field. 
These, in full generality, may include some parity-odd terms. Physically, they can appear
in models where the star formation 
physics is coupled to a parity odd sector, e.g.~models with an axion-photon coupling, or via some chiral gravity phenomena. As noted above, the coefficients of all parity-odd terms will be exactly zero in the absence of such effects. 

As is standard in the EFTofLSS, let us discuss various 
parity-violating terms in terms of their order 
in the loop expansion. Note that, unlike the parity-even operators, the loop expansion here does not match the derivative expansion, \textit{i.e.}\ the tree-level diagrams will actually be more suppressed than the loop ones in the gradient expansion. This happens because of 
additional constraints on the parity-odd terms due to the appearance of the Levi-Civita symbol, necessary for parity antisymmetry. 

At the tree level we have the following diagrams 
\beq
\qquad\raisebox{-0.0cm}{\includegraphicsbox[scale=0.57]{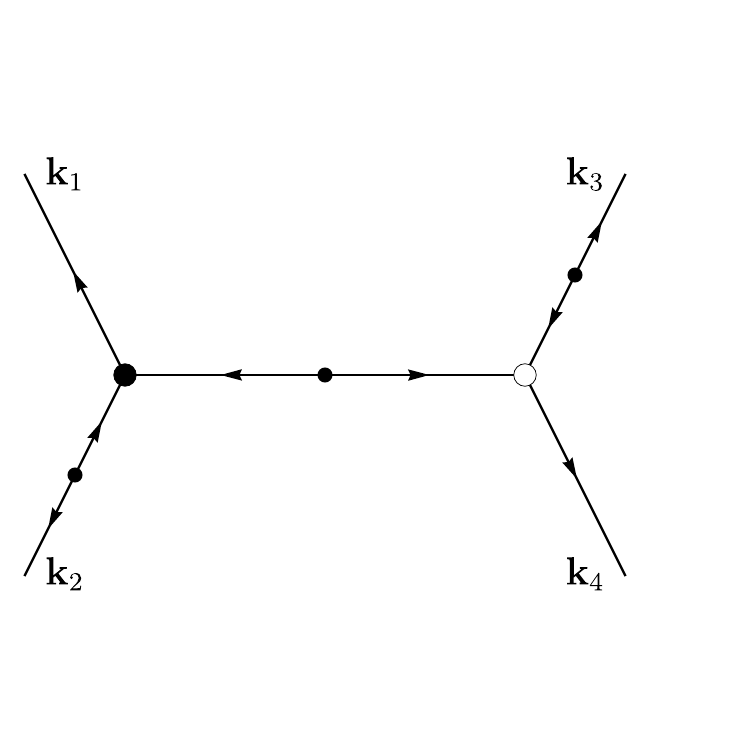}}\quad\text{and}\quad \raisebox{-0.0cm}{\includegraphicsbox[scale=0.57]{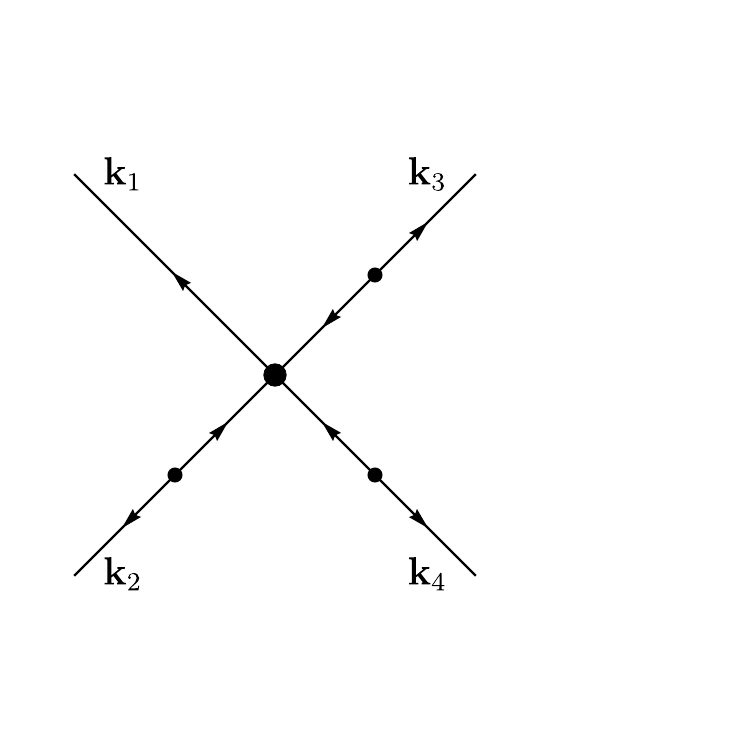}}
\eeq
which, in principle, may contribute to the parity-odd trispectrum. Here, the empty circle denotes a parity-even vertex and the filled circle a parity-odd one. The first of these two diagrams is zero at all orders in spatial derivatives since we have only the matter field $\delta$ to construct our operators, which is a scalar under rotations; as such, we cannot contract all the indices of the Levi-Civita symbol with only two powers of $\delta$ at our disposal. The second diagram is also zero unless one works at very high order in spatial derivatives: the leading operator with $n=3$ is given by
\beq
\label{higher_derivative_cubic_operator}
\delta_g\supset b^{(3)}_9 R^9_{\rm PO} \epsilon_{ijk}(\partial_i\delta)(\partial_j\partial^2\delta)(\partial_k\partial^4\delta)\,\,. 
\eeq 
Here we have factored out an overall $\smash{R^9_{\rm PO}}$ to make the bias coefficient $\smash{b^{(3)}_9}$ dimensionless (as indicated by the subscript). The scale $\smash{R_{\rm PO}}$ denotes the nonlocality scale of the new physics that generates these operators: given that it is related to parity-violating physics, this can be different to $R_\ast$ (which is usually taken to be of order of the Lagrangian radius of the halo, but see, e.g.,\,\citep{Schmidt:2017lqe,Cabass:2018hum} for discussions of scenarios where this is not the case). Below we will demonstrate that this contribution is highly suppressed in the power counting of the EFTofLSS, given the large number of spatial derivatives. This will imply that the leading contribution must come at one-loop order. 

Rotational invariance places strong constraints on one-loop diagrams, limiting us to 
\beq
\label{diagrams_EFTofLSS}
\qquad\raisebox{-0.0cm}{\includegraphicsbox[scale=0.57]{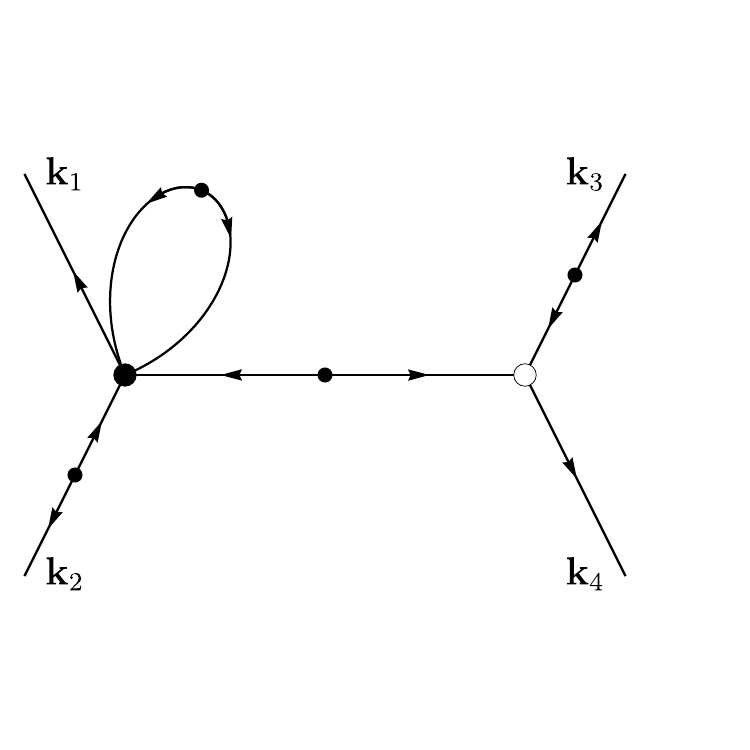}}\quad\text{and}\quad \raisebox{-0.0cm}{\includegraphicsbox[scale=0.57]{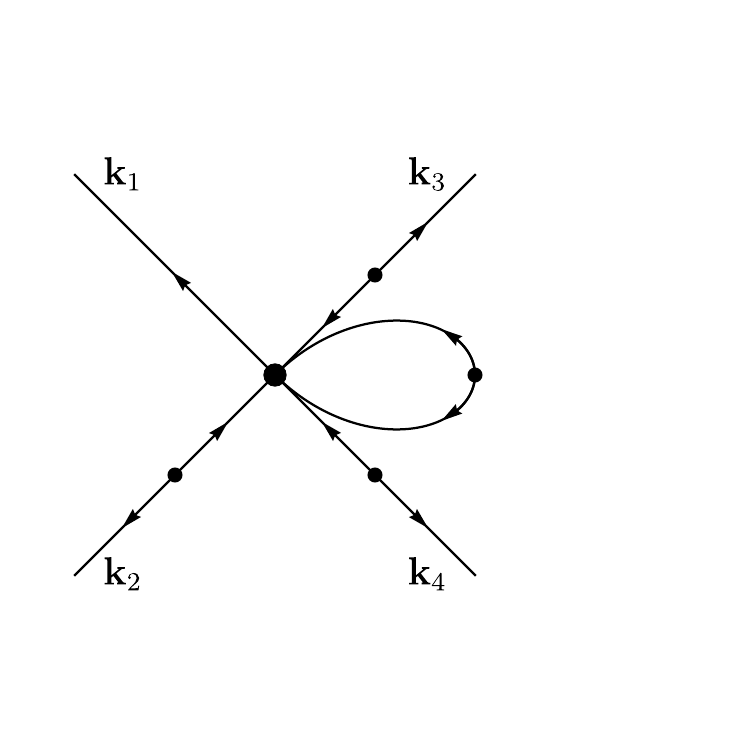}}\!\!\!\!\!\!\!\!\!\!\!\!\!\!\!\!\!\!\!\!.
\eeq
Let us focus on the first diagram. What is the form of the operator $\smash{{\cal O}^{(4)}}$ giving rise to the filled-circle vertex? At leading-order, this takes the form
\beq
\label{eft_operators}
\delta_g \supset 
b_1^{(4)} R_{\rm PO}
\epsilon_{ijk}(\Pi^{[1]}_{il})(\Pi^{[2]}_{lj})(\partial_k \delta)\,\,, 
\eeq 
while some examples of subleading operators are 
\beq
\label{eft_operators-subleading}
\delta_g \supset b_3^{(4)} R_{\rm PO}^3\epsilon_{ijk}  (\Pi^{[1]}_{il})(\Pi^{[2]}_{lj})(\partial_k\partial^2 \delta)+b_5^{(4)}R_{\rm PO}^5\delta\epsilon_{ijk}(\Pi^{[1]}_{il})(\partial_j \partial_l \partial^2 \delta)(\partial_k \delta)
\,\,. 
\eeq 
Here we follow \citep{Desjacques:2016bnm} and define $\smash{\Pi^{[1]}_{ij} \equiv (\partial_i\partial_j/\partial^2)\delta}$, with $\smash{\Pi^{[2]}_{ij}}$ being the element of the complete Eulerian basis of bias operators that starts at second order in perturbation theory, defined by (2.63) of \citep{Desjacques:2016bnm}. It is important to stress that, to generate a parity-violating signal, we must go beyond leading order in the expansion in spatial derivatives. This is clear from the expressions above. Indeed, the coefficient $\smash{b_{\cal O}^{(4)}}$ for the operator of \eqref{eft_operators} has dimension of $R_{\rm PO}$, while for the first and second operators of \eqref{eft_operators-subleading} it has dimension $R_{\rm PO}^3$ and $R_{\rm PO}^5$ respectively. 
We have again made this manifest by factoring out the powers of $R_{\rm PO}$ and defining the three dimensionless coefficients $\smash{b_1^{(4)}}$, $\smash{b_3^{(4)}}$ and $\smash{b_5^{(4)}}$. 

In absence of an hierarchy between relevant scales (\textit{i.e.}\ assuming $R_{\rm PO}\sim k_{\rm NL}^{-1}$), we expect: (1) that the first of the three operators listed above gives the largest signal, and (2) that this signal is smaller than even the (parity-preserving) gravitational trispectrum from the non-linear gravitational evolution at one loop order (which appears at zeroth order in $R_{\rm PO}$). Notice that the second diagram of \eqref{diagrams_EFTofLSS} enters at the same order in the power counting as we are always forced to have at least a spatial derivative $R_{\rm PO}\partial_i$ to contract the free index of the Levi-Civita symbol.

\subsection{Scaling Arguments}

\noindent We are now in the position to confirm that the one-loop diagram proportional to $\smash{b^{(4)}_1}$ is indeed the leading contribution. We can compare its size to the tree-level diagram coming from the higher-derivative operator of \eqref{higher_derivative_cubic_operator}. 
Focusing on the trispectrum in a configuration where the modes have all roughly the same size $k$ and approximating the linear matter power spectrum as a power law $\smash{(k/k_{\rm NL})^{n_\delta}/k_{\rm NL}^3}$ (where a spectral index $n_\delta$ close to ${-1.5}$ describes well the power spectrum at the scales used in our analysis) we find 
\beq
\frac{T_{\text{higher-derivative}}}{T_{\text{one-loop}}}\sim \frac{b_9^{(3)}}{b_1^{(4)}}\times(R_{\rm PO} k)^8\bigg(\frac{k}{k_{\rm NL}}\bigg)^{-{(3+n_\delta)}}\,\,. 
\eeq 
If we now assume a scaling universe \cite{Pajer:2013jj} we expect the nonlocality scale $R_{\rm PO}$ and the non-linear scale to be equal: then, we see that the tree-level diagram is strongly suppressed by $\smash{\sim(k/k_{\rm NL})^{6.5}}$, assuming similar magnitudes for the two bias coefficients. 

What can we instead conclude about the relative magnitude of the supposed parity-violating physics affecting galaxy formation and that from inflationary signals? We see that, unless the former has a spatial scale $R_{\rm PO}$ much shorter than the nonlinear scale or the halo Lagrangian radius $R_\ast$, its contributions to the galaxy four-point function is dominant with respect to inflationary signals. More precisely, we can estimate 
\beq
\frac{T_{\rm LSS}}{T_{\rm inflation}}\sim\dfrac{b_1^{(4)}}{\tau_{\rm NL}}\times\dfrac{(R_{\rm PO} k)\times\dfrac{1}{k^9_{\rm NL}}\bigg(\dfrac{k}{k_{\rm NL}}\bigg)^{3n_\delta}\times\bigg(\dfrac{k}{k_{\rm NL}}\bigg)^{3+n_\delta}}{(\Delta^2_\zeta)^3\dfrac{1}{k^9}\times\dfrac{1}{(\Delta^2_\zeta)^2}\bigg(\dfrac{k}{k_{\rm NL}}\bigg)^{2(3+n_\delta)}}\sim \dfrac{b_1^{(4)}}{\tau_{\rm NL}}\times\dfrac{(R_{\rm PO} k)\times\bigg(\dfrac{k}{k_{\rm NL}}\bigg)^{2(3+n_\delta)}}{\Delta^2_\zeta}\,\,, 
\eeq 
where we have assumed a scale-invariant ($\smash{\propto k^{-9}}$) trispectrum of the comoving curvature perturbation $\zeta$ and the factor of $\smash{(k/k_{\rm NL})^{2(3+n_\delta)}/(\Delta^2_\zeta)^2}$ at the denominator comes from four powers of the transfer function relating $\zeta$ to the linear matter density. 
The strong suppression in loops (scaling as $\smash{(k/k_{\rm NL})^{3}}$ for 
$n_\delta$ close to ${-1.5}$) and the additional suppression in spatial derivatives must be overcome for the late-time trispectrum to be larger than the inflationary contribution. Assuming again a scaling universe \cite{Pajer:2013jj} 
the suppression becomes $\smash{\sim(k/k_{\rm NL})^{4}}$. Taking $k_{\rm NL} = 0.5\Mpch$ at the redshift $z\simeq 0.5$ typical of our analysis, and taking a maximum momentum $k\simeq 0.1 \Mpch$, we see that 
\beq
\frac{T_{\rm LSS}}{T_{\rm inflation}}\sim10^5\times\frac{b_1^{(4)}}{\tau_{\rm NL}}\,\,. 
\eeq
Hence, for large values of $\tau_{\rm NL}\gtrsim 10^{5}$ (and assuming a dimensionless bias coefficient $\smash{b_1^{(4)}}$ of order unity), but still compatible with weak primordial non-Gaussianity $\smash{\tau_{\rm NL}\Delta^2_\zeta\ll1}$,\footnote{We remind the reader that in terms of the amplitudes $\smash{A^{(M_{\rm PO})}}$ and $\smash{A^{(\Lambda_{\rm PO}^2)}}$ of \S\ref{sec: results}, this constraint reads $\smash{A^{(M_{\rm PO})},A^{(\Lambda_{\rm PO}^2)}}\gtrsim 10^{-18}$.} we expect the inflationary parity violation to be dominant with respect to (possible) late-time contributions, even before cautioning that new physics at low redshifts is needed to source such effects. 

Before proceeding, let us also emphasize that so far we have focused on the EFTofLSS with \emph{Gaussian} initial conditions. In presence of some primordial (and parity-violating) non-Gaussianity one expects that additional operators are needed, in order to renormalize the contribution of short-wavevelength modes \cite{Assassi:2015fma}, akin to what happens for the scale-dependent bias in the case of local primordial non-Gaussianity.\footnote{Given that the Newtonian potential $\phi$ at the initial Lagrangian position and its derivatives would now be allowed in the bias expansion, we expect to be able to write operators that start at lower order in perturbations.} We leave the identification of these terms, and the estimation of their size, to future work.

\subsection{Velocity Contributions and Redshift-Space Distortions}

\noindent What happens once we consider the propagation of light from the galaxy rest frame to the observer? Parity-violating physics can affect the photon geodesics. However, in the non-relativistic regime $k/aH\gg 1$ typical of present and upcoming surveys, the only relevant projection effect is given by redshift-space distortions. These are the Doppler shift in the photon frequency due to the peculiar velocities of galaxies and only depend on the total intensity of photons emitted in the rest frame. Unless we include parity-violating operators in the bias expansion for the galaxy peculiar velocity, redshift-space distortions cannot lead to new contributions to the parity-odd trispectrum. 

In a similar manner to the above, one can ask what form the leading parity-odd operator in the EFTofLSS expansion of the galaxy velocity field $\textbf{v}_g$ should take. In this case, we need to construct an axial vector out of $\partial_i\delta$, the tidal field, and its higher-order generalizations $\smash{\Pi^{[n]}_{ij}}$ (we cannot use $\partial_i\phi$, where $\phi$ is the Newtonian potential, due to the equivalence principle). The leading non-vanishing operator consistent 
with symmetries of the 
EFTofLSS is then 
\beq
\label{rsd_operator}
\textbf{v}_g\supset b_{v}\epsilon_{ijk}\Pi^{[1]}_{jl}\Pi^{[2]}_{lk}\,\,, 
\eeq 
which we emphasize does not break the equivalence principle, since it is built out of local observables for an observer freely falling with the galaxies.\footnote{An equivalent way to see this is that both $\smash{\Pi^{[1]}_{ij}}$ and $\smash{\Pi^{[2]}_{ij}}$ are invariant under a shift of the Newtonian potential by a gradient mode.} 
The mapping from the rest-frame galaxy overdensity $\smash{\delta_g}$ to the redshift-space one $\smash{\tilde{\delta}_g}$ (following the notation of \citep{Desjacques:2016bnm}) contains the Kaiser term \citep{1987MNRAS.227....1K}
\beq
\label{final_rsd_equation}
\tilde{\delta}_g\supset\hat{n}_i\partial^i(\hat{n}_j v^j_g)\,\,, 
\eeq 
where $\smash{\hat{\textbf{n}}}$ is the line of sight. The contribution where $\mathbf{v}_g$ is given by \eqref{rsd_operator} is a cubic operator,\footnote{Indeed, unlike $\smash{\partial_i v^i_g}$ (which could appear in the galaxy density expansion), $\smash{\hat{n}_i v^i_g}$ starts at cubic order in perturbations.} and one without a suppression in spatial derivatives (indeed, $b_{v}$ is a dimensionless coefficient). Hence we expect that \eqref{rsd_operator} will contribute to the multipoles of the \emph{tree-level} trispectrum, provided there exists a mechanism to generate it. Unless $\smash{b^{(4)}_{1}}$ and $b_{v}$ are parametrically different, this ought to be the leading contribution from late-time parity violation in galaxy clustering in terms of the power counting of the EFTofLSS. However, it is important to keep in mind that higher-order multipoles of correlation functions are suppressed with respect to lower-order ones by powers of the growth rate $f$ divided by the linear bias $b_1$. It would be interesting to develop a pipeline to measure the multipoles of the redshift-space parity-odd galaxy trispectrum and put constraints on $\smash{b^{(4)}_{1}}$ and $b_{v}$: we leave this to future work.

\section{Conclusions} 
\label{sec: conclusion}

\noindent Does the early Universe conserve parity? In general, this is a difficult question to answer. Until we observe inflationary gravitational waves, we have no direct probe of parity-violation, thus must search instead for its signatures in scalar correlators, which appear only in the four-point function and beyond. Though the simplest models of inflation cannot source such a signal \citep{Cabass:2022rhr}, many non-standard theories can, involving, for example, a non-linear dispersion relation for the inflaton, the introduction of massive spinning fields, or time dependence of the couplings in the Lagrangian. In this work, we have considered several such models, using recent measurements of the galaxy 4PCF to place the first constraints on their amplitudes. 

In particular, we have considered Ghost Inflation, comprising a single field with a quadratic ($\omega\propto k^2$) dispersion relation, and the Cosmological Collider, whereupon the inflaton correlators are modulated due to the exchange of a massive spin-$1$ particle (for various choices of mass and sound speed). For each scenario, we forward-modeled the primordial correlator to obtain a late-time prediction for the 4PCF, and compared its form to the measurements presented in \citep{Philcox:2022hkh}. This yielded constraints on the couplings in the EFTI, all of which are consistent with zero (within $2\sigma$). Furthermore, these constraints were found to be parametrically relevant, in the context of the perturbativity bounds coming from the EFTI framework. 

The inflationary models considered herein are by no means exhaustive. For example, whilst Ghost Inflation provides a useful example of a non-linear dispersion relation, it suffers from a number of theoretical difficulties, as discussed in \citep{Cabass:2022rhr}.\footnote{More precisely, ghost condensation could conflict with black hole thermodynamics and the de Sitter entropy bound \cite{Dubovsky:2006vk,Arkani-Hamed:2007ryv}. However, it is also important to keep in mind that in Refs.~\cite{Mukohyama:2009rk,Mukohyama:2009um,Jazayeri:2016jav} it was shown that the subclass of Ghost inflation that is consistent with our Universe is not excluded by these theoretical constraints. We thank Shinji Mukohyama for discussions about this point.} If a prominent signal had been seen for this model, it would promote further investigation into other inflationary scenarios with similar dispersion relations. Secondly, we have considered only a single model for the Cosmological Collider; other couplings can exist (again see \citep{Cabass:2022rhr}), as well as models generated by an imbalance of the two helical exchange modes. One could further consider higher-order models, an example being the $U(1)$ gauge field couplings discussed in \citep{2016PhRvD..94h3503S} and constrained in \citep{Philcox:2022hkh}, or the exchange of chiral gravitational waves, produced via a Chern-Simons coupling. Given a primordial template, any model can be constrained following an (admittedly difficult) calculation similar to those in Appendix \ref{app: calculations}. It may additionally prove useful to define some general parametrization akin to $\{f_{\rm NL}^{\rm loc},f_{\rm NL}^{\rm eq},f_{\rm NL}^{\rm orth}\}$, onto which any primordial model can be projected.

The future will yield a vast increase in the volume of primordial modes surveyed. For LSS, the next generation of surveys will map out the distribution of around a hundred million galaxies, which should improve constraints on physical models such as the above by around an order of magnitude; these will strengthen still with proposed experiments such as MegaMapper \citep{Schlegel:2022vrv}. We additionally can make use of CMB data; the large-scale modes contained therein are predicted to be of great use in constraining primordial parity-violation (see, e.g., the forecasts of \citep{2016PhRvD..94h3503S}). However, the CMB is a two-dimensional field, and thus only parity sensitive on large scales, with statistically isotropic signals vanishing in the flat-sky regime (due to the identification of parity-reversal with a 3D rotation of the 2D CMB plane). In this sense, LSS data seems a more promising future avenue, though we caution that an experimental CMB parity-odd trispectrum study has yet to be performed. Finally, we note that many parity-odd inflationary models generate also a parity-even signature (such as in the Cosmological Collider). Often, the amplitudes of the two are related, and it is interesting to ask whether the physical models can be best constrained by parity-odd or parity-even measurements. For the CMB, the parity-even form is likely simpler (as it does not vanish in the small-scale limit), but for LSS, this observable is highly complex, due to the additional gravitational contributions (cf.\ \citep{Ivanov:2021kcd,Philcox:2022frc} for the bispectrum) which do not contribute source parity-odd trispectra. 

We close by commenting on the recent claims of a detection of parity-violation in LSS \citep{Philcox:2022hkh,Hou:2022wfj}. In this work, we have found that no evidence for inflationary parity-violation. Though our constraints are derived in the context of specific models, these templates are fairly generic, in that they are scale-independent, and span both exchange and contact diagrams, with a variety of microphysical parameters. Other models likely have significant cosines with the templates discussed herein, thus, to an extent, are already constrained. This lack of detection implies that the proposed signal of \citep{Philcox:2022hkh,Hou:2022wfj} contains a very different scale-dependence to that of inflation, which hints at a different resolution. As discussed in \S\ref{sec: late-time}, we can predict the form of late-time parity-violation using the EFTofLSS. Such contributions could arise from non-standard models of gravity (e.g., Chern-Simons gravity \citep{2009PhR...480....1A}) or hydrodynamics, and we find that they are generically suppressed on large scales, arising only from loop corrections or derivative operators. Again, this seems an unlikely explanation, given that the analysis was restricted to comparatively large scales ($r\gtrsim 20\Mpch$). As such, systematic effects, or poor understanding of the measurements' noise properties seems to be the most likely cause of the aforementioned detection, \resub{though we note that the space of possible cosmological explanations is large}. Whilst better data will help to understand the above, we stress that, if the signal is some unknown, and unsubtracted, systematic, its detection significance will only grow with the survey volume. Caution is warranted!

\acknowledgments

\noindent We thank Matias Zaldarriaga for 
enlightening conversations. 
We are additionally grateful to \resub{Matteo Biagetti}, Stephen Chen, Cyril Creque-Sarbinowski, Sadra Jazayeri, Marc Kamionkowski, \resub{Giorgio Orlando}, Enrico Pajer, Ue-Li Pen, Fabian Schmidt, David Stefanyszyn, and Yi Wang for insightful discussions. We also thank Fabian Schmidt for careful reading of the draft. 
GC acknowledges support from the Institute for Advanced Study. The work of MMI has been supported by NASA through the NASA Hubble Fellowship grant \#HST-HF2-51483.001-A awarded by the Space Telescope Science Institute, which is operated by the Association of Universities for Research in Astronomy, Incorporated, under NASA contract NAS5-26555. 
OHEP is a Junior Fellow of the Simons Society of Fellows and thanks the Institute for Advanced Study for their hospitality and abundance of baked goods. Part of this work was performed at the Aspen Center for Physics, which is supported by National Science Foundation grant PHY-1607611.

\appendix

\section{Computing 4PCF Templates}
\label{app: calculations}

\noindent In this appendix, we sketch the derivation of the galaxy 4PCF templates presented in \S\ref{sec: transform}. First, we note a number of general simplifications, before presenting specific results for the Ghost Inflation and Cosmological Collider templates.

\subsection{General Strategy}
\label{app-sub: tricks}

\noindent Starting from \eqref{eq: remapping}, we can simplify the 4PCF by shifting the permutation sum into the exponential term. This yields
\begin{equation}
\label{eq: remapping2}
\begin{split}
  \zeta_{\ell_1\ell_2\ell_3}(r_1,r_2,r_3) &= \int {\rm d}\hr_1{\rm d}\hr_2{\rm d}\hr_3\P^*_{\ell_1\ell_2\ell_3}(\hr_1,\hr_2,\hr_3)\left[\prod_{i=1}^4\int_{\vk_i}Z_1(\hk_i)M(k_i)\right]\delD{\vk_{1234}} \\ 
    &\;\;\;\;\,\times\tilde{T}(\vk_1,\vk_2,\vk_3,\vk_4)\sum_He^{i(\vk_1\cdot\vr_{H1}+\vk_2\cdot\vr_{H2}+\vk_3\cdot\vr_{H3}+\vk_4\cdot\vr_{H4})}\,\,,
\end{split}  
\end{equation}
where $\{H1,H2,H3,H4\}$ is one of the 24 permutations of $\{1,2,3,4\}$, $\vk_{1234}\equiv\vk_1+\vk_2+\vk_3+\vk_4$ and we have introduced $\vr_4=\vec 0$ for symmetry. Next, the integral over $\hr_i$ can be performed analytically, using the standard relation
\beq
	\int {\rm d}\hr\,e^{i\vk\cdot\vr}Y_{\ell m}(\hr) = 4\pi i^\ell j_\ell(kr)Y_{\ell m}(\hk) 
\eeq
for the spherical Bessel function $j_\ell(x)$. This allows us to write
\begin{equation}
\label{eq: remapping-simp}
\begin{split}
    \zeta_{\ell_1\ell_2\ell_3}(r_1,r_2,r_3) &= (4\pi)^{7/2}(-i)^{\ell_{123}}\sum_H\Phi_H\left[\prod_{i=1}^4\int_{\vk_i}Z_1(\hk_i)M(k_i)j_{\ell_{Hi}}(k_ir_{Hi})\right] \\
    &\;\;\;\;\,\times\P_{\ell_{H1}\ell_{H2}(\ell')\ell_{H3}\ell_{H4}}(\hk_1,\hk_2,\hk_3,\hk_4)\tilde{T}(\vk_1,\vk_2,\vk_3,\vk_4)\delD{\vk_{1234}}\,\,,
\end{split}  
\end{equation}
notating $\ell_{123}\equiv\ell_1+\ell_2+\ell_3$ and additionally inserting $1=\int {\rm d}\hr_4\,Y_{\ell_4m_4}(\hr_4)/\sqrt{4\pi}$ with $\ell_4=m_4=0$. \eqref{eq: remapping-simp} introduces a symmetry factor $\Phi_H\in\{\pm1\}$ (defined in \citep{Philcox:2022hkh}) and the four-coordinate basis function in $\hk$ \citep{2020arXiv201014418C}:
\begin{equation}
\label{eq: four-coord}
\begin{split}
    \P_{\ell_1\ell_2(\ell')\ell_3\ell_4}(\hk_1,\hk_2,\hk_3,\hk_4) &= (-1)^{\ell_{1234}}\sqrt{2\ell'+1}\sum_{m'}(-1)^{\ell'-m'}\sum_{m_1m_2m_3m_4}\tj{\ell_1}{\ell_2}{\ell'}{m_1}{m_2}{-m'}\tj{\ell'}{\ell_3}{\ell_4}{m'}{m_3}{m_4} \\
    &\;\;\;\;\,\times Y_{\ell_1m_1}(\hk_1)Y_{\ell_2m_2}(\hk_2)Y_{\ell_3m_3}(\hk_3)Y_{\ell_4m_4}(\hk_4)\,\,;
\end{split}  
\end{equation}
this is invariant under global rotations of $\hk_i$.

Another simplification concerns the perturbative kernels $Z_1$. Noting that
\begin{equation}
	Z_1(\hk) = 4\pi\sum_{\ell m}\left[\delta^{\rm K}_{\ell 0}\left(b+\frac{f}{3}\right)+\delta^{\rm K}_{\ell 2}\frac{2f}{15}\right]Y_{\ell m}(\hk)Y_{\ell m}^*(\hn) \equiv 4\pi\sum_{\ell m}Z_\ell Y_{\ell m}(\hk)Y_{\ell m}^*(\hn)\,\,,
\end{equation}
for line-of-sight $\hn$, we can average over $\hn$ by isotropy, which leads to
\begin{equation}
\begin{split}
	Z_1(\hk_1)Z_1(\hk_2)Z_1(\hk_3)Z_1(\hk_4) &\to (4\pi)^2\sum_{j_1j_2j_3j_4j'}\tjo{j_1}{j_2}{j'}\tjo{j'}{j_3}{j_4}Z_{j_1}Z_{j_2}Z_{j_3}Z_{j_4}\mathcal{C}_{j_1j_2j_3j_4j'} 
	\\
	&\;\;\;\;\;\;\times\P_{j_1j_2(j')j_3j_4}(\hk_1,\hk_2,\hk_3,\hk_4)
\end{split}
\end{equation} 
\citep{Philcox:2022hkh}, where $\mathcal{C}_{j_1\cdots j_n} = \sqrt{(2j_1+1)\cdots(2j_n+1)}$, and $j_i\in\{0,2\}$. 

Finally, the Dirac delta may be simplified in one of two ways. In the case of a \textit{contact} trispectrum (e.g., in Ghost Inflation), we may write
\beq
    \delD{\vk_{1234}} &=& \int {\rm d}\vx\,e^{i\vk_{1234}\cdot\vx} \nonumber \\ \nonumber
    &=&(4\pi)^4\sum_{L_1\cdots L_4M_1\cdots M_4}i^{L_{1234}}\int x^2 {\rm d}x\int {\rm d}\hx\,\left[\prod_{i=1}^4 \sum_{M_i}j_{L_i}(k_ix)Y_{L_iM_i}(\hk_i)Y^*_{L_iM_i}(\hx)\right] \\ 
    &=&(4\pi)^3\sum_{L_1\cdots L_4}(-i)^{L_{1234}}\int x^2 dx\,j_{L_1}(k_1x)j_{L_2}(k_2x)j_{L_3}(k_3x)j_{L_4}(k_4x)\\\nonumber
	&&\,\times\,\sum_{L'}\tjo{L_1}{L_2}{L'}\tjo{L'}{L_3}{L_4}\mathcal{C}_{L_1L_2L_3L_4L'}\P_{L_1L_2(L')L_3L_4}(\hk_1,\hk_2,\hk_3,\hk_4)\,\,,
\eeq
utilizing the plane-wave expansion in the second line, and the Gaunt integral and definitions of the four-particle basis function \eqref{eq: four-coord} in the third. For exchange trispectra, it is useful to instead introduce an internal (Mandelstam) momentum, $\vs\equiv\vk_1+\vk_2$. In this case;
\beq
    \delD{\vk_{1234}} &=& \int_{\vs }\delD{\vk_{12}+\vs}\delD{\vk_{34}-\vs} = \int_{\vs} {\rm d}\vx\,{\rm d}\vx'\,e^{i(\vk_{12}-\vs)\cdot\vx}e^{i(\vk_{34}+\vs)\cdot\vx'} \nonumber \\ \nonumber
    &=&(4\pi)^{5}\sum_{L_1\cdots L_6}i^{L_{1234}-L_5+L_6}\mathcal{C}_{L_1L_2L_3L_4L_5L_6}\tjo{L_1}{L_2}{L_5}\tjo{L_3}{L_4}{L_6}\\\nonumber
    &&\qquad\,\times\,\int_{\vs}\P_{L_1L_2L_5}(\hk_1,\hk_2,\hs)\P_{L_3L_4L_6}(\hk_3,\hk_4,\hs)\\
    &&\qquad\,\times\, \int_0^\infty x^2{\rm d}x\,j_{L_1}(k_1x)j_{L_2}(k_2x)j_{L_5}(sx)\int_0^\infty x'^2{\rm d}x'\,j_{L_3}(k_3x')j_{L_4}(k_4x')j_{L_6}(sx')\,\,.
\eeq
This can be simplified by integrating out $\hs$ (but retaining $s$, which appears also in the primordial trispectrum), leading to:
\beq\label{eq: dirac-exchange}
    \delD{\vk_{1234}} &=&(4\pi)^{4}\sum_{L_1\cdots L_4L'}(-i)^{L_{1234}}\mathcal{C}_{L_1L_2L_3L_4L'}\tjo{L_1}{L_2}{L'}\tjo{L_3}{L_4}{L'}\\\nonumber
    &&\qquad\,\times\, \int_0^\infty \frac{s^2{\rm d}s}{2\pi^2}\left[\int_0^\infty x^2{\rm d}x\,j_{L_1}(k_1x)j_{L_2}(k_2x)j_{L'}(sx)\right]\left[\int_0^\infty x'^2{\rm d}x'\,j_{L_3}(k_3x')j_{L_4}(k_4x')j_{L'}(sx')\right]\\\nonumber
    &&\qquad\,\times\,\P_{L_1L_2(L')L_3L_4}(\hk_1,\hk_2,\hk_3,\hk_4)\,\,.
\eeq

Though the above manipulations may seem only to add complexity, their benefit is that all the angular dependence is expressed purely in terms of basis functions, which (when a similar manipulation is performed for the inflationary trispectrum itself), can be straightforwardly combined and integrated over, leaving just a set of discrete summations and separable integrals in the radial components.

\subsection{Ghost Inflation}\label{app-sub: GI}

\noindent The primordial trispectra given in  \eqref{eq: primordial-ghost} can be separated into radial and angular coefficients. For the angular components, simplification is achieved utilizing the Cartesian forms of the isotropic basis functions \citep[cf.][]{Philcox:2022hkh,2020arXiv201014418C}. In particular, for the $\Lambda_{\rm PO}^2$ term we have
\begin{equation}
    \begin{split}
	\left(\hk_1\cdot\hk_2\times\hk_3\right)(\hk_1\cdot\hk_2) &= -i\frac{\sqrt{2}}{3\sqrt{3}}(4\pi)^{3}\P_{111}(\hk_1,\hk_2,\hk_3)\P_{110}(\hk_1,\hk_2,\hk_3) \\
	&= -\frac{i}{3}\sqrt\frac{2}{5}(4\pi)^2\P_{22(1)10}(\hk_1,\hk_2,\hk_3,\hk_4)\,\,,
    \end{split}
\end{equation}
contracting the two basis functions via \citep[\S6]{2020arXiv201014418C} to reach the second expression, which is written in terms of isotropic basis functions of four coordinates \eqref{eq: four-coord}. For the $M_{\rm PO}$ contribution the situation is less straightforward due to the large number of angles, but we eventually find
\beq
    \left(\hk_1\cdot\hk_2\times\hk_3\right)(\hk_2\cdot\hk_4)(\hk_1\cdot\hk_4)(\hk_2\cdot\hk_3) &=& -i\frac{\sqrt{2}}{9\sqrt{3}}(4\pi)^{8}\left[\P_{11(1)10}\P_{01(1)01}\P_{10(1)01}\P_{01(1)10}\right](\hk_1,\hk_2,\hk_3,\hk_4)\nonumber\\\nonumber
    &=&-i\frac{\sqrt{10}}{225}(4\pi)^2\left[\P_{01(1)22}-2\P_{03(3)22}+\frac{4\sqrt{2}}{5}\P_{21(1)22}+\P_{21(2)20}\right.\\
    &&\qquad-\sqrt{\frac{14}{5}}\P_{21(2)22}+\frac{2\sqrt{7}}{5}\P_{21(3)22}+\frac{\sqrt{3}}{5}\P_{23(1)22}-2\P_{23(2)20}\\\nonumber
    &&\qquad\left.-\sqrt{\frac{2}{35}}\P_{23(2)22}-\frac{2\sqrt{3}}{5}\P_{23(3)22}+\frac{6}{\sqrt{7}}\P_{23(4)22}\right](\hk_1,\hk_2,\hk_3,\hk_4)\\\nonumber
    &\equiv&-i(4\pi)^2\sum_{l_1l_2l_3l_4l'}c_{l_1l_2(l')l_3l_4}\P_{l_1l_2(l')l_3l_4}(\hk_1,\hk_2,\hk_3,\hk_4)\,\,,
\eeq
defining coefficients $c_{l_1l_2(l')l_3l_4}$ in the final line for brevity.

The radial piece is obtained by integrating the $\hk$-independent part of the trispectrum with respect to $k_i$, as in \eqref{eq: remapping-simp}. This is simplest to perform by switching the order of integration, placing the $\lambda$ integral (contained within $\cal T$ function of \eqref{eq: ghostT}) on the outside. For the operator proportional to $M_{\rm PO}^{-1}$, we find
\beq
	&&\left[\prod_{i=1}^4\int\frac{k_i^2{\rm d}k_i}{2\pi^2}M(k_i)j_{\ell_{Hi}}(k_ir_{Hi})j_{L_i}(k_ix)\right]k_1^{1/2}k_2^{3/2}k_3^{1/2}k_4^{1/2}\mathrm{Im}\mathcal{T}^{(11)}_{0,0,0,0}(k_1,k_2,k_3,k_4)\\\nonumber
	&&=\mathrm{Im}\int_{0}^{\infty} {\rm d}\lambda\, \lambda^{11}\,I_{3/4,1/2,\ell_{H1},L_1}(x,\lambda;r_{H1})I_{3/4,3/2,\ell_{H2},L_2}(x,\lambda;r_{H2})I_{3/4,1/2,\ell_{H3},L_3}(x,\lambda;r_{H3})I_{3/4,1/2,\ell_{H4},L_4}(x,\lambda;r_{H4})\,\,,
\eeq
and for that involving $\Lambda_{\rm PO}^{-2}$:
\beq
    &&\left[\prod_{i=1}^4\int\frac{k_i^2{\rm d}k_i}{2\pi^2}M(k_i)j_{\ell_{Hi}}(k_ir_{Hi})j_{L_i}(k_ix)\right]k_1^{1/2}k_2^{5/2}k_3^{3/2}k_4^{1/2}\mathcal{T}^{(13)}_{0,0,0,1}(k_1,k_2,k_3,k_4)\\\nonumber
    &&=\int_{0}^{\infty} {\rm d}\lambda\, \lambda^{13}\,I_{3/4,1/2,\ell_{H1},L_1}(x,\lambda;r_{H1})I_{3/4,5/2,\ell_{H2},L_2}(x,\lambda;r_{H2})I_{3/4,3/2,\ell_{H3},L_3}(x,\lambda;r_{H3})I_{-1/4,1/2,\ell_{H4},L_4}(x,\lambda;r_{H4})\,\,,
\eeq
defining
\beq\label{eq: ghostI}
	I_{\alpha,\beta,\ell,L}(x,\lambda;r) \equiv \int\frac{k^{2+\beta}{\rm d}k}{2\pi^2}M(k)j_\ell(kr)j_L(kx)H^{(1)}_{\alpha}(2ik^2\lambda^2)\,\,.
\eeq

In combination with the results of \S\ref{app-sub: tricks} we find the following forms for the ghost 4PCF:
\beq\label{eq: ghost-4pcf-tmp}
    \zeta^{(M_{\rm PO})}_{\ell_1\ell_2\ell_3}(r_1,r_2,r_3) &=&
    2(4\pi)^{19/2}(-i)^{\ell_{123}}\frac{\Lambda^5(H\tilde\Lambda)^{1/2}}{M_{\rm PO}\tilde\Lambda^5\Gamma(\frac{3}{4})^2}(\Delta_\zeta^2)^3\sum_H\Phi_H\sum_{L_1\cdots L_4L'}(-i)^{L_{1234}}\tjo{L_1}{L_2}{L'}\tjo{L'}{L_3}{L_4}\nonumber\\
    &&\,\times\,\mathcal{C}_{L_1L_2L_3L_4L'}\sum_{j_1j_2j_3j_4j'}\tjo{j_1}{j_2}{j'}\tjo{j'}{j_3}{j_4}Z_{j_1}Z_{j_2}Z_{j_3}Z_{j_4}\mathcal{C}_{j_1j_2j_3j_4j'}\\\nonumber
    &&\,\times\,\mathrm{Im}\int_0^\infty x^2 {\rm d}x\,\int_{0}^{\infty} {\rm d}\lambda\, \lambda^{11}I_{3/4,1/2,\ell_{H1},L_1}(x,\lambda;r_{H1})I_{3/4,3/2,\ell_{H2},L_2}(x,\lambda;r_{H2})\\\nonumber
    &&\qquad\qquad\,\times\,I_{3/4,1/2,\ell_{H3},L_3}(x,\lambda;r_{H3})I_{3/4,1/2,\ell_{H4},L_4}(x,\lambda;r_{H4})\\\nonumber
    &&\,\times\,\sum_{l_1\cdots l_4l'}c_{l_1l_2(l')l_3l_4}\int {\rm d}\hk_1{\rm d}\hk_2{\rm d}\hk_3{\rm d}\hk_4\left[\P_{l_1l_2(l')l_3l_4}\P_{L_1L_2(L')L_3L_4}\right.\\\nonumber
    &&\qquad\qquad\left.\,\times\,\P_{j_1j_2(j')j_3j_4}\P_{\ell_{H1}\ell_{H2}(\ell')\ell_{H3}\ell_{H4}}\right](\hk_1,\hk_2,\hk_3,\hk_4)\,\,,\\
    \zeta^{(\Lambda_{\rm PO}^2)}_{\ell_1\ell_2\ell_3}(r_1,r_2,r_3) &=& \frac{8\sqrt{2}}{3\sqrt{5}}(4\pi)^{19/2}(-i)^{\ell_{123}}\frac{\Lambda^5(H\tilde\Lambda)^{3/2}}{\Lambda^2_{\rm PO}\tilde\Lambda^6\Gamma(\tfrac{3}{4})^2}(\Delta_\zeta^2)^3\sum_H\Phi_H\sum_{L_1\cdots L_4L'}(-i)^{L_{1234}}\tjo{L_1}{L_2}{L'}\tjo{L'}{L_3}{L_4}\nonumber\\
    &&\,\times\,\mathcal{C}_{L_1L_2L_3L_4L'}\sum_{j_1j_2j_3j_4j'}\tjo{j_1}{j_2}{j'}\tjo{j'}{j_3}{j_4}Z_{j_1}Z_{j_2}Z_{j_3}Z_{j_4}\mathcal{C}_{j_1j_2j_3j_4j'}\\\nonumber
    &&\times\,\int_0^\infty x^2{\rm d}x\,\int_0^\infty {\rm d}\lambda\,\lambda^{13}I_{3/4,1/2,\ell_{H1},L_1}(x,\lambda;r_{H1})I_{3/4,5/2,\ell_{H2},L_2}(x,\lambda;r_{H2})\\\nonumber
    &&\qquad\qquad\qquad\qquad\times\,I_{3/4,3/2,\ell_{H3},L_3}(x,\lambda;r_{H3})I_{-1/4,1/2,\ell_{H4},L_4}(x,\lambda;r_{H4})\\\nonumber
    &&\,\times\,\int {\rm d}\hk_1{\rm d}\hk_2{\rm d}\hk_3{\rm d}\hk_4\left[\P_{22(1)10}\P_{L_1L_2(L')L_3L_4}\P_{j_1j_2(j')j_3j_4}\P_{\ell_{H1}\ell_{H2}(\ell')\ell_{H3}\ell_{H4}}\right](\hk_1,\hk_2,\hk_3,\hk_4)\,\,.
\eeq
The final line of each expression involves the integral over four sets of basis functions; these can be evaluated in terms of $9j$ symbols \citep{2020arXiv201014418C}, and written in terms of angular coupling matrices, given by
\beq\label{eq: ghost-coupling}
	\mathcal{M}_{L_1L_2(L')L_3L_4}^{\ell_{H1}\ell_{H2}(\ell')\ell_{H3}\ell_{H4}} &=&\mathcal{C}_{\ell_{H1}\ell_{H2}\ell'\ell_{H3}\ell_{H4}}\mathcal{C}_{L_1L_2L'L_3L_4}\sum_{l_1l_2l_3l_4l'}c_{l_1l_2(l')l_3l_4}\mathcal{C}_{l_1l_2(l')l_3l_4}\\\nonumber
	&&\,\times\,\sum_{j_1j_2j_3j_4j'}\tjo{j_1}{j_2}{j'}\tjo{j'}{j_3}{j_4}Z_{j_1}Z_{j_2}Z_{j_3}Z_{j_4}\mathcal{C}^2_{j_1j_2j_3j_4j'}\\\nonumber
	&&\,\times\,\sum_{\lambda_1\lambda_2\lambda_{12}\lambda_3}(-1)^{\lambda_1+\lambda_2+\lambda_3+\lambda_4}\mathcal{C}_{\lambda_1\lambda_2\lambda_{12}\lambda_3\lambda_4}^2\tjo{l_1}{L_1}{\lambda_1}\tjo{l_2}{L_2}{\lambda_2}\tjo{l_3}{L_3}{\lambda_3}\tjo{l_4}{L_4}{\lambda_4}\\\nonumber
	&&\,\times\,\tjo{j_1}{\ell_{H1}}{\lambda_1}\tjo{j_2}{\ell_{H2}}{\lambda_2}\tjo{j_3}{\ell_{H3}}{\lambda_3}\tjo{j_4}{\ell_{H4}}{L_4}\\\nonumber
	&&\,\times\,\begin{Bmatrix}l_1&l_2&l'\\L_1&L_2&L'\\\lambda_1&\lambda_2&\lambda_{12}\end{Bmatrix}\begin{Bmatrix}l'&l_3&l_4\\L'&L_3&L_4\\\lambda_{12}&\lambda_3&L_4\end{Bmatrix}\begin{Bmatrix}j_1&j_2&j'\\\ell_{H1}&\ell_{H2}&\ell'\\\lambda_1&\lambda_2&\lambda_{12}\end{Bmatrix}\begin{Bmatrix}j'&j_3&j_4\\\ell'&\ell_{H3}&\ell_{H4}\\\lambda_{12}&\lambda_3&L_4\end{Bmatrix}
\eeq
and
\beq
	\mathcal{N}_{L_1L_2(L')L_3L_4}^{\ell_{H1}\ell_{H2}(\ell')\ell_{H3}\ell_{H4}} &=&15\mathcal{C}_{\ell_{H1}\ell_{H2}\ell'\ell_{H3}\ell_{H4}}\mathcal{C}_{L_1L_2L'L_3L_4}\sqrt{2L_4+1}\sum_{j_1j_2j_3j_4j'}\tjo{j_1}{j_2}{j'}\tjo{j'}{j_3}{j_4}Z_{j_1}Z_{j_2}Z_{j_3}Z_{j_4}	\nonumber
\\\nonumber
	&&\,\times\,\mathcal{C}^2_{j_1j_2j_3j_4j'}\sum_{\lambda_1\lambda_2\lambda_{12}\lambda_3}(-1)^{\lambda_1+\lambda_2+\lambda_3}\mathcal{C}_{\lambda_1\lambda_2\lambda_{12}\lambda_3}^2\tjo{2}{L_1}{\lambda_1}\tjo{2}{L_2}{\lambda_2}\tjo{1}{L_3}{\lambda_3}\\\nonumber
	&&\,\times\,\tjo{j_1}{\ell_{H1}}{\lambda_1}\tjo{j_2}{\ell_{H2}}{\lambda_2}\tjo{j_3}{\ell_{H3}}{\lambda_3}\tjo{j_4}{\ell_{H4}}{L_4}\\
	&&\,\times\,\begin{Bmatrix}2&2&1\\L_1&L_2&L'\\\lambda_1&\lambda_2&\lambda_{12}\end{Bmatrix}\begin{Bmatrix}1&1&0\\L'&L_3&L_4\\\lambda_{12}&\lambda_3&L_4\end{Bmatrix}\begin{Bmatrix}j_1&j_2&j'\\\ell_{H1}&\ell_{H2}&\ell'\\\lambda_1&\lambda_2&\lambda_{12}\end{Bmatrix}\begin{Bmatrix}j'&j_3&j_4\\\ell'&\ell_{H3}&\ell_{H4}\\\lambda_{12}&\lambda_3&L_4\end{Bmatrix}\,\,,
\eeq
where the curly parentheses are $9j$ symbols. We additionally note that $L_1+L_2+L_3+L_4$ is even (from the $3j$ symbols) and $\ell_1+\ell_2+\ell_3$ is odd, thus the expression is purely imaginary (using the properties of Hankel functions). Inserting these into \eqref{eq: ghost-4pcf-tmp} leads to the final expressions given in \eqref{eq: ghost-4pcf-I}\,\&\,\eqref{eq: ghost-4pcf-II}.

\subsection{Cosmological Collider}\label{app-sub: CC}

\noindent To evaluate the Cosmological Collider template, it is convenient to first split the primordial correlator of \eqref{eq: primordial-collider-oliver} into two pieces joined by an angular factor:
\beq
    \tilde{T}_{\lambda_1\lambda_3}(\vk_1,\vk_2,\vk_3,\vk_4)&=&-ic_s^4\frac{\lambda_1\lambda_3}{\resub{2}H}(\Delta_\zeta^2)^4\sin \pi\left(\nu+\frac{1}{2}\right)t^A(k_1,k_2,s)t^B(k_3,k_4,s)\\\nonumber
    &&\qquad\qquad\,\times\,(\hk_1\cdot\hk_2)(\hk_3\cdot\hk_4)\left(\hk_2\cdot(\hk_3\times\hk_4)\right)\,\,,
\eeq
using the definitions
\beq
    t^A(k_1,k_2,s) &=& k_1^{-2}k_2^{-1}(k_1-k_2)[k_{12}J_{3}(c_{s}k_{12},s)+ c_{s}k_{1}k_{2}J_{4}(c_{s} k_{12},s)]\\\nonumber
    t^B(k_3,k_4,s) &=& k_3^{-1}k_4^{-1}(k_3-k_4)[k_{34}J_{4}(c_{s}k_{34},s)+c_{s}k_{3}k_{4}J_{5}(c_{s}k_{34},s)]\,\,.
\eeq
As before, this contains a cross-product and is purely imaginary. Unlike for Ghost Inflation, this is an exchange diagram, thus has dependence on the exchange momentum $s$.

To simplify the angular component, we use the basis functions of \eqref{eq: four-coord}, writing:
\begin{equation}
    \begin{split}
    \left(\hk_2\cdot(\hk_3\times\hk_4)\right)(\hk_1\cdot\hk_2)(\hk_3\cdot\hk_4) &= i\frac{\sqrt{2}}{\resub{9}}(4\pi)^{7/2}\P_{111}(\hk_2,\hk_3,\hk_4)\P_{11(0)11}(\hk_1,\hk_2,\hk_3,\hk_4)\\ 
    &= i\frac{(4\pi)^2}{9\sqrt{5}}\left[2\,\P_{12(1)22}(\hk_1,\hk_2,\hk_3,\hk_4)-\sqrt{2}\,\P_{10(1)22}(\hk_1,\hk_2,\hk_3,\hk_4)\right]\,\,.
    \end{split}
\end{equation}
For the radial integrals, we first rewrite the momentum-conserving delta function in terms of \eqref{eq: dirac-exchange}; this will lead to radial integrals of the form:
\beq\label{eq: colliderQ}
    Q^{\ell_1\ell_2,X}_{L_1L_2L'}(s;r_1,r_2) = \int_0^\infty x^2{\rm d}x\,j_{L'}(sx)\prod_{i=1}^2\left[\int\frac{k_i^2{\rm d}k_i}{2\pi^2}M(k_i)j_{\ell_{i}}(k_ir_i)j_{L_i}(k_ix)\right]t^X(k_1,k_2,s)\,\,,
\eeq
for $X\in\{A,B\}$.

Utilizing the results of Appendix \ref{app-sub: tricks}, we find that the 4PCF can be written
\beq
    \zeta^{(\lambda_1\lambda_3)}_{\ell_1\ell_2\ell_3}(r_1,r_2,r_3) &=& (4\pi)^{15/2}(-i)^{\ell_{123}}\frac{c_s^4\lambda_1\lambda_3}{\resub{18}\sqrt{5}H}(\Delta_\zeta^2)^4\sin\pi\left(\nu+\frac{1}{2}\right)\sum_H\Phi_H\sum_{L_1\ldots L_4L'}(-i)^{L_{1234}}\tjo{L_1}{L_2}{L'}\tjo{L'}{L_3}{L_4}\nonumber\\
    &&\,\times\,\mathcal{C}_{L_1\ldots L_4L'}\sum_{j_1j_2j_3j_4j'}\tjo{j_1}{j_2}{j'}\tjo{j'}{j_3}{j_4}Z_{j_1}Z_{j_2}Z_{j_3}Z_{j_4}\mathcal{C}_{j_1j_2j_3j_4j'}\\\nonumber
    &&\,\times\,\int\frac{s^2{\rm d}s}{2\pi^2}Q^{\ell_{H1}\ell_{H2},A}_{L_1L_2L'}(s;r_{H1},r_{H2})Q^{\ell_{H3}\ell_{H4},B}_{L_3L_4L'}(s;r_{H3},r_{H4})\\\nonumber
    &&\,\times\,\int {\rm d}\hk_1{\rm d}\hk_2{\rm d}\hk_3{\rm d}\hk_4\,\left[(2\,\P_{12(1)22}-\sqrt{2}\,\P_{10(1)22})\P_{j_1j_2(j')j_3j_4}\right.\\\nonumber
    &&\qquad\qquad\qquad\qquad\qquad\,\left.\times\,\P_{L_1L_2(L')L_3L_4}\P_{\ell_{H1}\ell_{H2}(\ell')\ell_{H3}\ell_{H4}}\right](\hk_1,\hk_2,\hk_3,\hk_4)\,\,.
\eeq
This may be further decomposed by defining the coupling matrix (integrating over basis functions to obtain Wigner $9j$ symbols, as before):
\beq\label{eq: couplingO}
    \mathcal{O}_{L_1L_2(L')L_3L_4}^{\ell_{H1}\ell_{H2}(\ell')\ell_{H3}\ell_{H4}} 
    &=& 15\,\mathcal{C}_{\ell_{H1}\ell_{H2}\ell'\ell_{H3}\ell_{H4}}\mathcal{C}_{L_1L_2L'L_3L_4}\sum_{j_1j_2j_3j_4j'}\tjo{j_1}{j_2}{j'}\tjo{j'}{j_3}{j_4}Z_{j_1}Z_{j_2}Z_{j_3}Z_{j_4}\\\nonumber
    &&\times\,\mathcal{C}^2_{j_1j_2j_3j_4j'}\sum_{\lambda_1\lambda_2\lambda_{12}\lambda_3\lambda_4}(-1)^{\lambda_{1234}}\mathcal{C}^2_{\lambda_1\lambda_2\lambda_{12}\lambda_3\lambda_4}\tjo{1}{L_1}{\lambda_1}\tjo{2}{L_3}{\lambda_3}\tjo{2}{L_4}{\lambda_4}\\\nonumber
    &&\times\,\tjo{j_1}{\ell_{H1}}{\lambda_1}\tjo{j_2}{\ell_{H2}}{\lambda_2}\tjo{j_3}{\ell_{H3}}{\lambda_3}\tjo{j_4}{\ell_{H4}}{\lambda_4}\begin{Bmatrix}j_1 & j_2 & j'\\ \ell_{H1} & \ell_{H2} & \ell'\\\lambda_{1} & \lambda_2 & \lambda_{12}\end{Bmatrix}\begin{Bmatrix}j' & j_3 & j_4\\ \ell' & \ell_{H3} & \ell_{H4}\\\lambda_{12} & \lambda_3 & \lambda_4\end{Bmatrix}\\\nonumber
    &&\times\,\left[2\sqrt{5}\tjo{2}{L_2}{\lambda_2}\begin{Bmatrix}1 & 2 & 1\\ L_1 & L_2 & L'\\\lambda_1 & \lambda_2 & \lambda_{12}\end{Bmatrix}-\sqrt{2}\tjo{0}{L_2}{\lambda_2}\begin{Bmatrix}1 & 0 & 1\\ L_1 & L_2 & L'\\\lambda_1 & \lambda_2 & \lambda_{12}\end{Bmatrix}\right]
    \begin{Bmatrix}1 & 2 & 2\\ L' & L_3 & L_4\\\lambda_{12} & \lambda_3 & \lambda_4\end{Bmatrix}\,\,,
\eeq
yielding the final expression given in \eqref{eq: CC-4pcf}.

\bibliographystyle{apsrev4-1}
\bibliography{refs}

\end{document}